\documentclass[11pt]{article}

\usepackage{hyperref}
\usepackage{bm}
\usepackage{booktabs}
\usepackage[margin=15pt,font=small,labelfont={bf,sf},justification=justified]{caption}
\usepackage[superscript,biblabel]{cite}
\usepackage{color}
\usepackage{float}
\usepackage{placeins}
\usepackage{graphicx}
\usepackage{lscape}
\usepackage[bbgreekl]{mathbbol}
\usepackage{mathtools}
\usepackage{multibib}
\usepackage{multirow}
\usepackage{tabularx}
\usepackage{titleref}
\usepackage{url}
\usepackage{siunitx}
\usepackage{subfigure}
\usepackage[export]{adjustbox}
\usepackage{epsfig}

\newif\ifbembo
\newif\ifcharter
\newif\iferewhon
\newif\iflibertine
\newif\iflibertinealt
\newif\ifpalantino
\newif\iftimesnewroman

\bembofalse
\chartertrue
\erewhonfalse
\libertinefalse
\libertinealtfalse
\palantinofalse
\timesnewromanfalse

\ifbembo
  \usepackage[p,osf]{ETbb} 
  \usepackage[scaled=.95,type1]{cabin} 
  \usepackage[varqu,varl]{zi4}
  \usepackage[T1]{fontenc}
  \usepackage[libertine,vvarbb]{newtxmath}
  \usepackage[cal=boondoxo,bb=boondox]{mathalfa}
\fi

\ifcharter
  \usepackage[scaled=.98,p]{XCharter}
  \usepackage[scaled=1.04,varqu,varl]{inconsolata}
  \usepackage[type1]{cabin}
  \usepackage[uprightscript,xcharter,vvarbb,scaled=1.05]{newtxmath}
  \usepackage[cal=euler,scr=boondoxo,bb=boondox]{mathalpha}  
\fi

\iferewhon
  \usepackage[p,osf,scaled=.98,space]{erewhon} 
  \usepackage[varqu,varl]{inconsolata} 
  \usepackage[type1,scaled=.95]{cabin} 
  \usepackage[utopia,vvarbb]{newtxmath}
\fi

\iflibertine
  \usepackage[sb]{libertinus}
  \usepackage[T1]{fontenc}
  \usepackage{textcomp}
  \usepackage[varqu,varl]{zi4}
  \usepackage{libertinust1math} 
  \usepackage[scr=boondoxo,bb=boondox]{mathalpha} 
\fi

\iflibertinealt
  \usepackage[sb]{libertinus}
  \usepackage[T1]{fontenc}
  \usepackage{textcomp}
  \usepackage{libertinust1math} 
  \usepackage[scr=boondoxo,bb=boondox]{mathalpha} 
\fi

\ifpalantino
  \usepackage[largesc]{newpxtext}
  \usepackage[vvarbb]{newpxmath}
\fi

\iftimesnewroman
  \usepackage{newtxtext} 
  \usepackage[scaled=0.95]{inconsolata} 
  \usepackage[vvarbb]{newtxmath} 
\fi

\usepackage{microtype}
\usepackage{titling}
\usepackage{authblk}
\pretitle{\begin{center}\bfseries\Large}
\posttitle{\end{center}}

\predate{\begin{center}\normalsize}
\postdate{\end{center}}

\usepackage{titling}
\usepackage{authblk} 
\pretitle{\begin{center}\bfseries\sffamily\LARGE}
\posttitle{\end{center}}
\usepackage{abstract}

\allowdisplaybreaks
\usepackage{sectsty} 
\allsectionsfont{\sffamily}

\usepackage[left=0.5in,right=0.5in,top=0.5in,bottom=0.5in,includefoot,heightrounded]{geometry}

\makeatletter
\patchcmd{\LS@rot}{90}{-90}{}{}
\patchcmd{\endlandscape}{90}{-90}{}{}
\makeatother

\title{Benchmarking the Immersed Boundary Method for Viscoelastic Flows}
\author[1]{Cole Gruninger}
\author[2]{Aaron Barrett}
\author[1]{Fuhui Fang}
\author[1,3--5]{M. Gregory Forest}
\author[1,3--7,*]{Boyce E. Griffith}
\affil[1]{Department of Mathematics, University North Carolina, Chapel Hill, NC, USA}
\affil[2]{Department of Mathematics, University of Utah, Salt Lake City, UT, USA}
\affil[3]{Department of Applied Physical Sciences, University of North Carolina, Chapel Hill, NC, USA}
\affil[4]{Department of Biomedical Engineering, University of North Carolina, Chaple Hill, NC, USA}
\affil[5]{Carolina Center for Interdisciplinary Applied Mathematics, University of North Carolina, Chapel Hill, NC, USA}
\affil[6]{Computational Medicine Program, University of North Carolina School of Medicine, Chapel Hill, NC, USA}
\affil[7]{McAllister Heart Institute, University of North Carolina School of Medicine, Chapel Hill, NC, USA\vspace{0.5\baselineskip}}
\affil[*]{To whom correspondence should be addressed; email: \texttt{boyceg@email.unc.edu}}

\newcommand{\totStress}{\bbsigma}
\newcommand{\fluidStress}{\bbsigma_\text{f}}
\newcommand{\polymerStress}{\bbsigma_\text{p}}
\newcommand{\conformTens}{\mathbb{C}}
\newcommand{\velocity}{\textbf{u}}
\newcommand{\coordMap}{\bm{\chi}}
\newcommand{\lagVel}{\textbf{U}}
\newcommand{\eulForce}{\textbf{f}}
\newcommand{\lagForce}{\textbf{F}}
\newcommand{\density}{\rho}
\newcommand{\viscosity}{\mu}
\newcommand{\fluidViscosity}{\mu_\text{s}}
\newcommand{\polymerViscosity}{\mu_\text{p}}
\newcommand{\relaxationTime}{\lambda_\text{r}}
\newcommand{\reptationTime}{\lambda_\text{d}}
\newcommand{\rouseTime}{\lambda^{}_\text{R}}
\newcommand{\relaxationFunc}{\textbf{g}}
\renewcommand{\Re}{\text{Re}}
\newcommand{\Wi}{\text{Wi}}

\newcommand{\tran}{^{\mkern-1.5mu\mathsf{T}}}
\newcommand{\domain}{\Omega}
\newcommand{\interface}{\Gamma_t}
\newcommand{\interfaceInitial}{\Gamma_0}
\newcommand{\xx}{\mathbf{x}}
\newcommand{\XX}{\mathbf{X}}
\newcommand{\parens}[1]{\mathopen{}\left(#1\right)\mathclose{}}
\newcommand{\parenstwo}[1]{\left(#1\right)}
\newcommand{\parensExp}[1]{\left(#1\right)}
\newcommand{\grad}{\nabla}
\newcommand{\II}{\mathbb{I}}
\newcommand{\tr}[1]{\text{tr}\parens{#1}}

\newcommand{\spread}{\mathbfcal{S}}
\newcommand{\interp}{\mathbfcal{I}}

\begin{document}
\maketitle

\begin{abstract}
\noindent We present and analyze a series of benchmark tests regarding the application of the immersed boundary (IB) method to viscoelastic flows through and around non-trivial, stationary geometries. The IB method is widely used for the simulation of biological fluid dynamics and other modeling scenarios where a structure is immersed in a fluid. Although the IB method has been most commonly used to model systems with viscous incompressible fluids, it also can be applied to visoelastic fluids, and has enabled the study of a wide variety of dynamical problems including the settling of vesicles and the swimming of elastic filaments in fluids modeled by the Oldroyd-B constuitive equation. However, to date, relatively little work has explored the accuracy or convergence properties of the numerical scheme. Herein, we present benchmarking results for an IB solver applied to viscoelastic flows in and around non-trivial geometries using the idealized Oldroyd-B and more realistic, polymer-entanglement-based Rolie-Poly constitutive equations. We use two-dimensional numerical test cases along with results from rheology experiments to benchmark the IB method and compare it to more complex finite element and finite volume viscoelastic flow solvers. Additionally, we analyze different choices of regularized delta function and relative Lagrangian grid spacings which allow us to identify and recommend the key choices of these numerical parameters depending on the present flow regime. 
\end{abstract}

\section{Introduction}
The immersed boundary (IB) method is a mathematical formulation and numerical scheme utilized to model the presence of a structure immersed in a fluid \cite{peskin2002,griffith2020}. The IB method was originally developed by Peskin to model the flow of blood within the heart \cite{peskin1972,peskin1977}. The IB method has since been adapted to study a variety of fluid-structure interaction (FSI) problems spanning applications from a broad array of fields, including cardiac mechanics \cite{griffith_luo_2009,hasan2017,griffith_heart2012,chen2016,crowl2011}, platelet adhesion and aggregation \cite{skorczewski2014,fogelson2008}, insect flight \cite{jones2015,Santhanakrishnan2018}, and undulatory swimming \cite{alben2013,bhalla2014,tytell2014,bale2015,hoover2017,nangia2017}. The IB method employs a combination of Lagrangian and Eulerian variables that interact with one another through the action of integral transforms with the Dirac delta function. In the IB formulation, the velocity and incompressibility of the fluid-structure system are described using Eulerian variables, and the configurations of the immersed structures and their resultant forces are described using Lagrangian variables. The Eulerian variables are discretized on a structured Cartesian grid and the Lagrangian variables are discretized on a curvelinear mesh that moves freely through the Cartesian grid without needing to conform to it. Interactions between the Eulerian and Lagrangian variables is mediated through discretized integral transforms against a regularized Dirac delta kernel function which are used both to interpolate the discrete Eulerian velocity onto the Lagrangian curvelinear mesh and also to spread forces from the Lagrangian curvelinear mesh onto the Cartesian grid. 

Since its inception, the IB method has typically used a Newtonian description of the fluid in which the stress tensor associated with the fluid is linearly related to the fluid's velocity gradient. Fluids that are comprised of small, relatively inert molecules can be accurately modeled using a Newtonian description; examples are water, liquid nitrogen, and liquid argon. The success of the Newtonian description hinges on the assumption that, even for turbulent flows, intermolecular distances between the neighboring molecules that make up the fluid are relatively preserved, so that the energy dissipation mechanism, modeled by viscosity in a Newtonian fluid, is constant \cite{morozov2015}. However, for fluids whose molecular constituents are perturbed by the flow, the energy dissipation mechanism becomes a function of the local flow dynamics, and the Newtonian model subsequently breaks down. Fluids for which the Newtonian model fails to provide an adequate description are typically called ``complex'' or ``non-Newtonian'' fluids. Non-Newtonian fluids are typically comprised of large polymeric molecules (chains) that, either individually at dilute concentrations, or at the scale of entanglements at higher concentrations, can be stretched out of their equilibrium configurations, leading to modes of storage and relaxation of stresses in response to gradients present in the velocity field. Consequently, many non-Newtonian fluids display simultaneously the behaviours of a viscous liquid and an elastic solid. Such fluids are appropriately termed ``viscoelastic'' fluids. 

The majority of theoretical and empirical studies examining the accuracy of the IB method focus on the evolution of viscous fluids. The use of regularized delta functions in the IB method generally limits its accuracy to first order, a limitation which has been corroborated by several previous studies \cite{beyer1992,li1994,lai2000,griffith2005,mori2008}. The intuitive explanation as to why the IB method is generally first-order accurate is that the fluid pressure and normal derivative of the fluid velocity have jump discontinuities across the immersed structure that are smoothed out by the regularized delta function. As a result, fluid stresses generally do not converge pointwise at the immersed boundary. Nonetheless, in practice it is observed that the velocity does converge pointwise at a first-order rate at the boundary. This finding was first proved for the case of an incompressible Newtonian fluid by Mori in the context of Stokes flow \cite{mori2008}. Mori also provided a heuristic analysis that concluded that, so long as one refines the Lagrangian mesh spacing proportional to the Eulerian mesh spacing, one should expect the velocity to converge at a second-order rate in the $L^1$ grid norm, a three-halves rate in the $L^2$ norm, and a first-order rate in the $L^\infty$ norm. This same $L^2$ error estimate was proven more recently by Heltai and Lei in the context of a finite element immersed boundary method applied to a simple elliptic equation, for which a priori estimates are readily accessible \cite{heltai2020}. Subsequent theoretical and empirical work in the context of Stokes flow by Liu and Mori have further analyzed how the accuracy of the velocity is affected by the choice of the regularized delta function employed \cite{liu2012}. Liu and Mori's error estimates demonstrated that smoother regularized delta functions are able to remove presence of spurious oscillations or Gibbs type phenomena present in the computed solution. In support of Liu and Mori's finding, Yang et al. \cite{yang2009} introduced smoothed versions of some of the regularized delta functions derived by Peskin \cite{peskin2002,roma1999}, and demonstrated empirically that they were better at removing spurious oscillations present in the Lagrangian forces defined along the immersed boundary. More recently, in the context of the immersed finite element/difference (IFED) method, Lee and Griffith \cite{lee2022} analyzed how the choice of regularized delta function and relative mesh spacing of the Lagrangian and Eulerian grid affects the accuracy of computed Newtonian flows. Lee and Griffith concluded that, so long as pressure loads on the immersed structure are not substantial, using regularized delta functions with narrower supports coupled with a relatively coarse discretization of the Lagrangian structure with respect to the background Eulerian grid provides a substantial increase in accuracy of the IFED method.

Although the IB method has been mostly used in conjunction with Newtonian fluid models, there has been some previous work with viscoelastic fluid models \cite{kim2018,salazar2016,chrispell2011,teran2010}. These studies utilized fluid descriptions based on the Oldroyd-B model of a viscoelastic fluid, which incorporates an additional viscoelastic stress tensor that evolves with the fluid \cite{morozov2015}. It remains unclear whether the IB method can reliably simulate viscoelastic FSI problems because the polymeric stress tensor evolution directly depends on the velocity gradient. Nonetheless, prior work using the IB method to simulate viscoleastic FSI has demonstrated reasonable results. For instance, Ma et al.\cite{ma2020} \ developed an immersed boundary method with a lattice-Boltzmann fluid solver and were able to achieve convergent velocities, even at high Weissenberg numbers. Zhu et al.\cite{zhu2014,zhu2017,zhu2019} has also empirically demonstrated the utility and accuracy of the IB method coupled with a lattice-Boltzman fluid solver for both the fluid velocity and evolution of the polymeric stress. However, Zhu et al. did not consider the pointwise accuracy associated with the computed polymeric stress like we do here. Stein et al. \cite{stein2017,stein2019} investigated the IB method applied to Oldroyd-B fluid models and found that although the IB method does not produce a pointwise convergent viscoelastic stress at interfaces, the method does tend to provide convergent net forces on immersed structures, despite the net forces being inaccurate for under resolved grids.

Herein, we extend the investigations by Stein et al. further and demonstrate that the IB method used in conjunction with a viscoelastic fluid model obtains pointwise first-order accurate fluid velocities, first and half-order rates of convergence of the viscoelastic stress in the $L^1$ and $L^2$ grid norms respectively, and effectively replicates macrorheology experiments. Furthermore, we find that although stresses do not converge pointwise up to the immersed boundary, net forces can be computed with first order accuracy by summing over the Lagrangian forces. The accuracy of computed net forces depend substantially on the regularized delta function and the relative Lagrangian grid spacings employed. Findings for the convergence properties of net-forces on structures are compared against the effect the regularized delta function and relative Lagrangian grid spacings have on the computed fluid velocity and viscoelastic stresses whose accuracies appear to depend on the identity of the regularized delta function, but not the relative Lagrangian grid spacings unless pressure loads on the immersed structure are significant.

\section{Model Description}\label{sec:model_description}
We consider an infinitesimally thin interface $\interfaceInitial$ immersed in an incompressible fluid in a domain $\domain$. Our formulation uses both Eulerian coordinates $\xx \in \Omega$ and Lagrangian reference coordinates $\XX \in \Gamma_0$. The fluid is described by a velocity $\velocity\parens{\xx,t}$ and a Cauchy stress tensor $\totStress\parens{\xx,t}$. The configuration of the interface at time $t$ is given by the mapping $\coordMap\parens{\XX,t}\in\interface$. The equations of motion are
\begin{align}
    \density\parens{\frac{\partial\velocity\parens{\xx,t}}{\partial t} + \velocity\parens{\xx,t}\cdot\grad\velocity\parens{\xx,t}} &= \grad\cdot\totStress\parens{\xx,t} + \eulForce\parens{\xx,t}, \label{eq:mom}\\
    \grad\cdot\velocity\parens{\xx,t} &= 0,   \label{eq:mass}\\
    \eulForce\parens{\xx,t} &= \int_{\interfaceInitial}\lagForce\parens{\XX,t}\,\delta\parens{\xx-\coordMap\parens{\XX,t}}\,\text{d}A, \label{eq:eulForce}\\
    \frac{\partial\coordMap\parens{\XX,t}}{\partial t} &= \lagVel\parens{\XX,t} = \int_{\domain}\velocity\parens{\xx,t}\,\delta\parens{\xx - \coordMap\parens{\XX,t}}\,\text{d}\xx, \label{eq:interfaceMotion}
\end{align}
in which $\density$ is the density of the fluid, $\text{d}A$ is the surface element associated with $\Gamma_0$, $\eulForce\parens{\xx,t}$ is the Eulerian body force density, $\lagForce\parens{\XX,t}$ is the Lagrangian force density, and $\lagVel\parens{\XX,t}$ is the Lagrangian velocity. Equations \eqref{eq:mom} and \eqref{eq:mass} describe the conservation of momentum and the incompressibility constraint on the velocity field, respectively. Equation \eqref{eq:interfaceMotion} specifies that the immersed boundary moves according to the local fluid velocity. The Lagrangian force density $\lagForce\parens{\XX,t}$ generated by that movement induces an Eulerian force density $\eulForce\parens{\xx,t}$ through equation \eqref{eq:eulForce}. The integral transforms against the Dirac delta function present in equations \eqref{eq:eulForce} and \eqref{eq:interfaceMotion} are the heart of the mathematical formulation of the IB method and establish the FSI coupling between the immersed boundary and the fluid. In this study, we focus on the case in which the position of the immersed boundary is fixed, for which the interface force density $\lagForce\parens{\XX,t}$ is a Lagrange multiplier enforcing the rigidity of the interface. Although it is possible to devise IB methods that determine the exact Lagrange multiplier force density $\lagForce\parens{\XX,t}$ \cite{kallemov2016,usabiaga2016}, we instead approximate the Lagrange multiplier force density by a penalty spring force that opposes the motion of the immersed boundary

\begin{equation}\label{eq:penalty_spring}
    \lagForce\parens{\XX,t} = \kappa\parens{\coordMap\parens{\XX,0} - \coordMap\parens{\XX,t}},
\end{equation}
in which $\kappa$ is a penalty stiffness parameter. We use this penalty force formulation to avoid solving the poorly conditioned extended saddle-point system imposed by using an exact Lagrange multiplier formulation. In practice, we choose the stiffness parameter $\kappa$ to be approximately the largest stable value permitted by the numerical scheme so the Lagrangian markers used to discretize the structure move a distance no more than $\frac{h}{2}$, in which $h$ is the uniform Cartesian grid increment.  

We focus on viscoelastic fluid models in which the total stress $\totStress\parens{\xx,t}$ can be decomposed into a Newtonian solvent and polymeric stress,
\begin{equation}
    \totStress\parens{\xx,t} = \fluidStress\parens{\xx,t} + \polymerStress\parens{\xx,t}.
\end{equation}
The Newtonian stress is
\begin{equation}
\fluidStress\parens{\xx,t} = -p\parens{\xx,t}\II + \frac{\fluidViscosity}{2}\parens{\grad\velocity\parens{\xx,t} + \grad\velocity\parens{\xx,t}\tran},
\end{equation}
in which $p\parens{\xx,t}$ is the isotropic pressure and $\fluidViscosity$ is the solvent contribution to the viscosity. The viscoelastic fluid models we consider herein can be derived from a microscopic bead-spring model, in which the polymeric stress is linearly related to the conformation tensor by 
\begin{equation}\label{eq:stress_conformation_relation}
\polymerStress\parens{\xx,t} = \frac{\polymerViscosity}{\relaxationTime}\parens{\conformTens\parens{\xx,t} - \II},
\end{equation}
in which $\polymerViscosity$ is the polymeric contribution to the viscosity, $\relaxationTime$ is the relaxation time of the fluid, and $\conformTens\parens{\xx,t}$ is the conformation tensor. The conformation tensor represents an ensemble average over the dyad formed by the local end-to-end vectors characterizing the bead-spring polymers \cite{morozov2015}. The evolution of the conformation tensor satisfies 
\begin{equation}\label{eq:conformTens}
    \overset{\triangledown}{\conformTens}= \relaxationFunc\parens{\conformTens\parens{\xx,t}},
\end{equation}
in which $\relaxationFunc\parens{\conformTens}$ is the model dependent relaxation function of the polymer configurations, and $\overset{\triangledown}{\conformTens}$ is James Oldroyd's frame-invariant upper-convected time derivative\cite{oldroyd1950}, which is defined by
\begin{equation}\label{eq:upper_convective_t_derivative}
    \overset{\triangledown}{\conformTens} = \frac{\partial \conformTens}{\partial t} + \velocity\cdot\grad\conformTens - \grad\velocity\cdot\conformTens - \conformTens\cdot\grad\velocity\tran.
\end{equation}
We consider two models, the Oldroyd-B model
\begin{equation}
\relaxationFunc\parens{\conformTens} = -\frac{1}{\relaxationTime}\parens{\conformTens - \II},
\end{equation}
and the Rolie-Poly model
\begin{equation}
\relaxationFunc\parens{\conformTens} = -\frac{1}{\reptationTime}\parens{\conformTens - \II} - \frac{2}{\rouseTime}\parens{1 - \sqrt{\frac{2}{\tr{\conformTens}}}}\parens{\conformTens + \beta\parensExp{\frac{\tr{\conformTens}}{2}}^\delta\parens{\conformTens-\II}},
\end{equation}
in which $\reptationTime$ and $\rouseTime$ are the repatation and Rouse times, respectively, and $\tr{\conformTens}$ is the trace of the conformation tensor. In the context of the Rolie-Poly model, we take the relaxation time $\relaxationTime$ in equation \eqref{eq:stress_conformation_relation} to be the reptation time. The parameters $\beta$ and $\delta$ are chosen empirically through a fitting procedure.

Viscoelastic flow is often characterized using the dimensionless Reynolds $\Re = \frac{\density U L}{\viscosity}$ and Weissenberg $\Wi = \frac{U L}{\relaxationTime}$ numbers, in which $U$ is a characteristic flow speed and $L$ is a characteristic length scale. The Reynolds number is a ratio between the viscous and inertial forces, and in the applications considered herein, we focus on the small Reynolds number regime relevant for experiments and biological applications of interest. The Weissenberg number is the ratio between the strain rate and the relaxation time of the polymers and represents the ratio of elastic to viscous forces. 
\section{Numerical Implementation}\label{numerics}
This section describes the spatial and temporal discretization of the variables and differential operators involved in the IB formulation for viscoelastic fluids described in Section \ref{sec:model_description}. The spatial discretization is described in two spatial dimensions, and its extension to three spatial dimensions is straightforward. We conclude this section by referencing the software employed in this study and by discussing the appropriate scaling of stiffness parameter $\kappa$ in equation \eqref{penalty_spring_scaling} to maintain stability of the discrete system under spatio-temporal grid refinement.

\subsection{Spatial Discretization}\label{sec:space_disc}
\begin{figure}[t!]
    \centering
    \includegraphics[width=.55\linewidth]{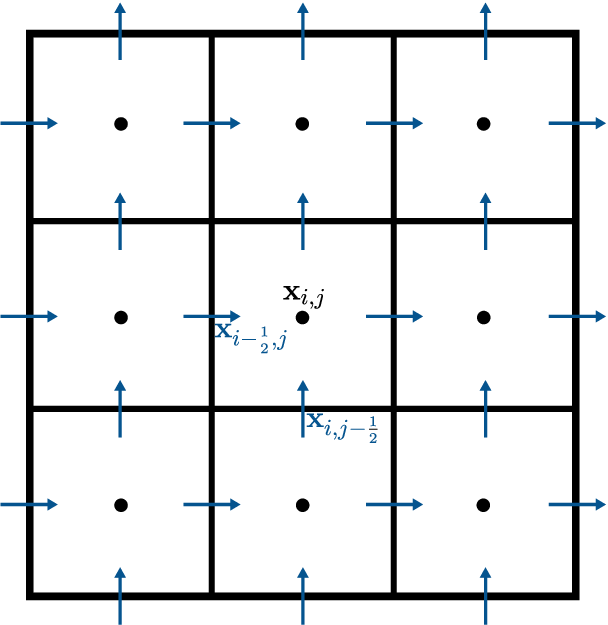}
    \caption{Locations of cell-centered and face-centered quantities about Cartesian grid cell $\mathbf{x}_{i,j}$.}
    \label{fig:macgrid}
\end{figure}

 The physical domain $\Omega$ is described by a fixed $N\times N$ Cartesian grid with uniform meshwidths $\Delta x = \Delta y = h = \frac{L}{N} $, in which $L$ is some chosen characteristic length scale. The Eulerian variables are discretized on a staggered-grid in which the components of the velocity are associated with the grid cell faces and the pressure and conformation tensor degrees of freedom are associated with the grid cell centers. See Figure \ref{fig:macgrid}.
 
We denote centers of the Cartesian grid cells are located at the points $\mathbf{x}_{i,j} = \left(\left( i + \frac{1}{2}\right)h,\left(j + \frac{1}{2}\right)h\right)$, for $i,j = 0,\ldots,N-1$. The centers of the $x$-edges of the grid cells are the points $\mathbf{x}_{i-\frac{1}{2},j} = \left(ih,\left(j+\frac{1}{2}\right)h\right)$, where $i = 0,\ldots,N$ and $j = 0,\ldots,N-1$. The centers of the $y$-edges are the points $\mathbf{x}_{i,j-\frac{1}{2}} = \left(\left(i + \frac{1}{2}\right)h,jh\right)$, for $i = 0,\ldots,N-1$ and $j = 0,\ldots,N$. The pressure $p(\mathbf{x},t)$ and conformation tensor $\conformTens$ are defined at the centers of the Cartesian grid cells. The horizontal component, $u$, of the velocity is defined on the $x$-edges of the grid cells and the vertical component of the velocity, $v$, is defined along the $y$-edges of the grid cells. The horizontal component and vertical components of the force are likewise defined along the $x$-edges and $y$-edges of the grid cells, respectively. 

We now detail the discretization of the spatial differential operators present in the Eulerian equations of motion. Standard compact second-order finite differences are used for the divergence and Laplacian of the velocity and the gradient of the velocity \cite{griffith2009}. The Laplacian of the velocity $\grad_h^2\velocity$ and the pressure gradient $\grad_h p$ are evaluated at cell faces, and the divergence of the velocity $\grad_h\cdot\velocity$ is evaluated at cell centers. The convective term $\velocity\cdot\nabla_h\velocity$ in the momentum equation \eqref{eq:mom} is approximated using the xsPPM7 scheme\cite{rider2007}, a variant of the piecewise parabolic method\cite{colella1984}. The convective term $\mathbf{u}\cdot\nabla_h\conformTens$ contained in the definition of the upper-convective time derivative \eqref{eq:upper_convective_t_derivative} is approximated using the second-order wave propagation algorithm described by Ketchenson et al.\cite{ketcheson2013} 

To approximate the velocity gradient at cell centers, we use the following finite difference stencils
\begin{align}
    \parenstwo{\frac{\partial u}{\partial x}}_{i,j} &\approx \frac{u_{i+\frac{1}{2},j}-u_{i-\frac{1}{2},j}}{h}, \\
    \parenstwo{\frac{\partial u}{\partial y}}_{i,j} &\approx \frac{u_{i+\frac{1}{2},j+1} + u_{i-\frac{1}{2},j+1} - u_{i+\frac{1}{2},j-1} - u_{i-\frac{1}{2},j-1}}{4h}, \\
    \parenstwo{\frac{\partial v}{\partial x}}_{i,j} &\approx \frac{v_{i+1,j+\frac{1}{2}} + v_{i+1,j-\frac{1}{2}} - v_{i-1,j+\frac{1}{2}} - v_{i-1,j-\frac{1}{2}}}{4h}, \\
    \parenstwo{\frac{\partial v}{\partial y}}_{i,j} &\approx \frac{v_{i,j+\frac{1}{2}}-v_{i,j-\frac{1}{2}}}{h}.
\end{align}
In practice, we substitute the divergence of the polymeric stress from equation \eqref{eq:mom} with the divergence of the conformation tensor, as per equation \eqref{eq:stress_conformation_relation}. To approximate the divergence of the conformation tensor along cell edges, we use
\begin{align}
    \parenstwo{\frac{\partial C_{xx}}{\partial x}}_{i-\frac{1}{2},j} &\approx \frac{C_{xx_{i,j}} - C_{xx_{i-1,j}}}{h}, \\
    \parenstwo{\frac{\partial C_{xy}}{\partial y}}_{i-\frac{1}{2},j} &\approx \frac{C_{xy_{i+1,j+1}} + C_{xy_{i-1,j+1}} - C_{xy_{i+1,j-1}} - C_{xy_{i-1,j-1}}}{4h}, \\
    \parenstwo{\frac{\partial C_{yx}}{\partial x}}_{i,j-\frac{1}{2}} &\approx \frac{C_{yx_{i+1,j+1}} + C_{yx_{i+1,j-1}} - C_{yx_{i-1,j+1}} - C_{yx_{i-1,j-1}}}{4h}, \\
    \parenstwo{\frac{\partial C_{yy}}{\partial y}}_{i,j-\frac{1}{2}} &\approx \frac{C_{yy_{i,j}} - C_{yy_{i,j-1}}}{h}.
\end{align}
We remark that the stencils above represent a centered, formally second-order accurate approximation of both the velocity gradient and divergence of the conformation tensor. 

\subsection{Lagrangian grid spacing}\label{sec:lag_grid}
The reference configuration $\Gamma_0$ of each immersed boundary we consider in section \ref{numerical tests} is made up of a finite collection of smooth curves. Each curve comprising $\Gamma_0$ can be expressed by a parametrization $\mathbf{X}(s)$ that we discretize by placing a finite number $M$ of Lagrangian markers located at $\{\XX(s_k)\}_{k=1}^{k=M}$ along the curve. We choose the positions of the Lagrangian markers so that they are approximately equally spaced along $\Gamma_0$. To do so, we ensure that the arc length increment,
\begin{align}\label{eq:arclengthincrement}
    \Delta X_{k} = \int_{s_{k}}^{s_{k+1}} \left|\left|\frac{\text{d}\mathbf{X}_k(s^{\prime})}{\text{d}s^{\prime}}\right|\right|_2\,\text{d}s^{\prime}
\end{align}
remains approximately equal to an a priori chosen idealized arc length increment $\Delta X$. Generally, the total arc length $L$ of a curve $\XX(s)$ making up $\Gamma_0$ is not an integer multiple of the ideal arc length increment $\Delta X$. In this situation, if the curve parameterized by $\mathbf{X}(s)$ is closed, we first compute $M = \lfloor \frac{L}{\Delta X}\rceil$, in which $\lfloor \cdot \rceil$ is the function that rounds to the nearest integer, and set $\Delta X = \frac{L}{M}$. We then place Lagrangian markers along $\mathbf{X}(s)$ starting at $\mathbf{X}(s_1)$ and then at $\mathbf{X}(s_{k+1})$ by solving 
\begin{align}\label{eqn:lag_dist_eq}
    \int_{s_{k}}^{s_{k+1}} \left|\left| \frac{\text{d}\mathbf{X}(s^{\prime})}{\text{d}s^{\prime}}\right|\right|\text{d}s^{\prime} - \Delta X = 0,
\end{align}
for $s_{k+1}$ and $k = 1,\ldots,M-1$. In the case in which the curve $\mathbf{X}(s)$ is open, we still set $M = \lfloor \frac{L}{\Delta X}\rceil$, but choose the ideal arc length increment to be $\Delta X= \frac{L}{M - 1}$. We then follow the same procedure and place a Lagrangian marker starting at $\mathbf{X}(s_{k_1})$ and solve equation \eqref{eqn:lag_dist_eq} for the remaining locations of the Lagrangian markers $\mathbf{X}(s_{k+1})$ for $k = 1,\ldots,M- 1$. 
With our discretization procedure in place, we next define the mesh factor, $M_{\text{fac}}$, to be the ratio of the idealized arc length increment $\Delta X$ to the finest Eulerian grid mesh width $h$,
\begin{align}
    M_{\text{fac}} = \frac{\Delta X}{h}.
\end{align}
\begin{figure}[t!]
    \centering
    \includegraphics[scale = 0.6]{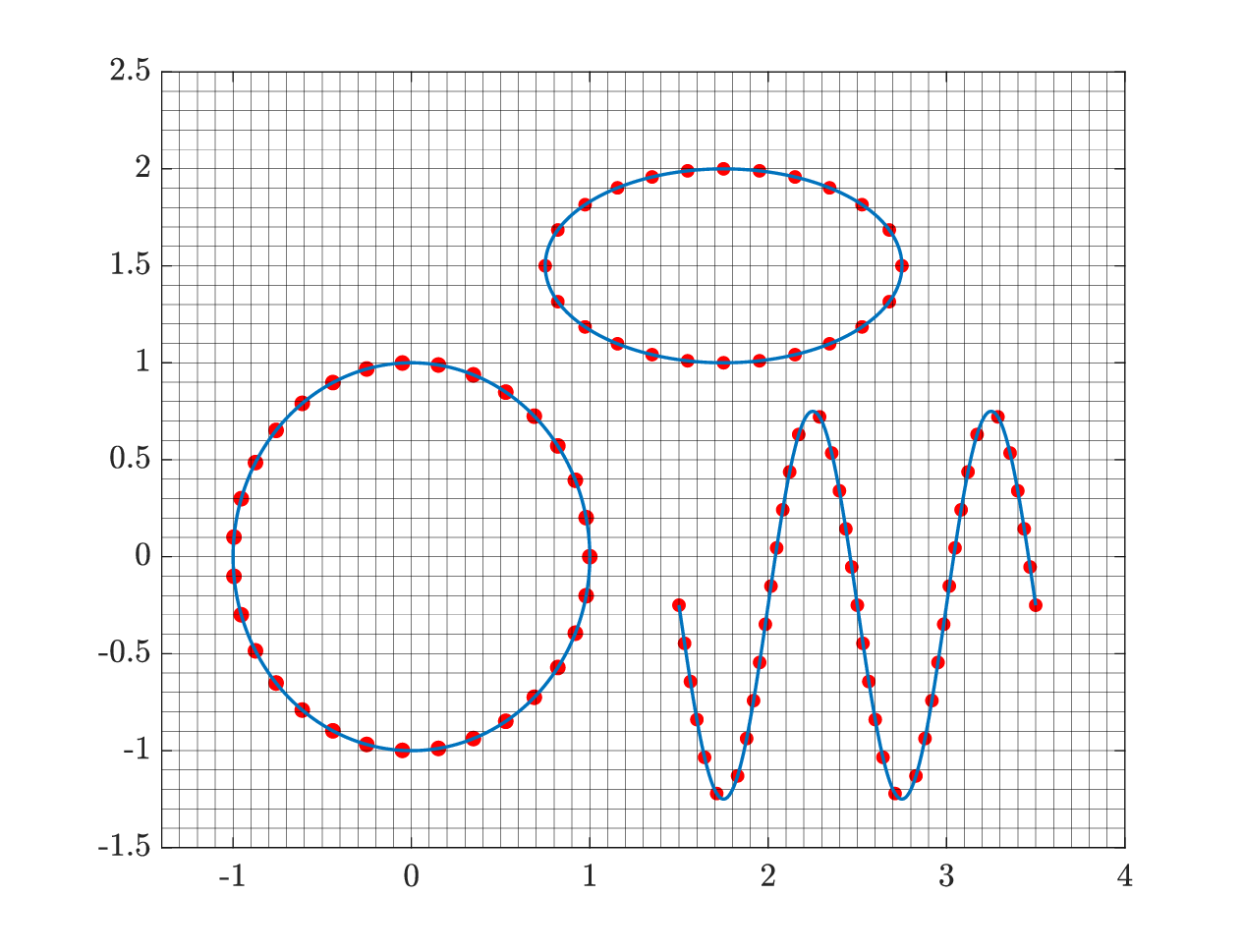}
    \caption{Example Lagrangian discretization of a circle, ellipse, and a sine curve. Here $h = 0.1$ and the mesh factor is $M_{\text{fac}} = \frac{\Delta X}{h} = 2$, in which $\Delta X$ is the ideal arc length increment separating each Lagrangian marker along a curve. Since the total arc length of each curve is not necessarily an integer multiple of the Cartesian grid increment, the total number of Lagrangian markers placed along each curve is obtained by rounding the ratio of the total arc length of a curve to the ideal arc length increment to the nearest integer. When the arc length of a given curve is not an explicitly invertible function, Lagrangian markers are placed along the curve by approximately inverting the arc length function through the use of a nonlinear solver with an error tolerance specified to be at least three orders of magnitude smaller than the idealized arc length increment.} 
    \label{fig:example_discretization}
\end{figure}
To construct the initial configuration of the immersed boundary $\Gamma_0$, we choose the ideal arc length increment $\Delta X$ based on a particular choice of $M_{\text{fac}}$. To illustrate, suppose we set $M_{\text{fac}} = 2$ and choose $\Gamma_0$ to be made up of a circle of radius $r = 1$ and with center $(0,0)$, an ellipse with center $(1.75,1.5)$, width $2$, and height $1$, and a sine curve of frequency $2\pi$ and mean amplitude $-0.25$ starting at $x = 1.5$ and ending at $x = 3.5$. The paramterizations of these curves are $\mathbf{X}_1 = (\cos(s_1),\sin(s_1))$ with $s_1 \in [0,2\pi)$, $\mathbf{X}_2 = (1.75 + \cos(s_2),1.5 + \sin(s_1))$ with $s_2\in[0,2\pi)$, and $\mathbf{X}_3 = \left(s_3,-0.25 + \sin(2\pi s_2)\right)$ with $s_3\in[1.5,3.5]$, respectively. Following our discretization procedure above, we produce the discretization of $\Gamma_0$ shown in Figure \ref{fig:example_discretization}.

\subsection{Spreading and Interpolation}\label{sec:interp}
The interaction between the fluid and structure is mediated by the equations \eqref{eq:eulForce} and \eqref{eq:interfaceMotion}. In the IB method, the singular delta functions supported on the immersed boundary are replaced by regularized delta functions $\delta_h(x)$. Given the Lagrangian force $\lagForce\parens{\XX,t} = (F_1(\XX,t),F_2(\XX,t))$, the forces $\eulForce\parens{\xx,t} = \left(f_1(\xx,t),f_2(\xx,t)\right)$ on the Cartesian grid are computed as
\begin{align}
    \parens{f_1}_{i-\frac{1}{2},j} &= \sum_{I = 1}^M F_1\parens{\XX_I,t}\delta_h\parens{\xx_{i-\frac{1}{2},j} - \coordMap\parens{\XX_I,t}}\Delta \XX_I, \\
    \parens{f_2}_{i,j-\frac{1}{2}} &= \sum_{I = 1}^M F_2\parens{\XX_I,t}\delta_h\parens{\xx_{i,j-\frac{1}{2}} - \coordMap\parens{\XX_I,t}}\Delta \XX_I,
\end{align}
in which $M$ is the total number of Lagrangian points and $\Delta \XX_I$ is the arc-length element associated with Lagrangian grid point $\XX_I$, which is described in section \ref{sec:lag_grid}. We denote the discrete spreading operation of the force onto the background grid as
\begin{equation}
    \eulForce = \spread\left[\coordMap\right]\lagForce,
\end{equation}
in which $\spread$ is the discrete force spreading operator. The horizontal $U_k$ and vertical $V_k$ components of the velocity associated with the $k^{\text{th}}$ Lagrangian marker point are computed via 
\begin{align}
  U_k = h^2\sum_{i,j=0}^{N-1}u_{i-\frac{1}{2},j}\,\delta_h\parens{\xx_{i-\frac{1}{2},j}-\coordMap\parens{\XX_k,t}}, \\
  V_k = h^2\sum_{i,j=0}^{N-1}v_{i,j-\frac{1}{2}}\,\delta_h\parens{\xx_{i,j-\frac{1}{2}} - \coordMap\parens{\XX_k,t}},
\end{align}
which is a discrete approximation to the integral equation given in equation \eqref{eq:interfaceMotion}.
To keep our notation compact, we denote the discretized interpolation operation of the velocity onto the Lagrangian structure as
\begin{equation}
    \lagVel = \interp\left[\coordMap\right]\velocity,
\end{equation}
in which $\interp$ is the discrete interpolation operator. We construct the interpolation operator by requiring that it be adjoint to the spreading operator, $\interp = \spread^*$. This requirement is fulfilled by using the same regularized delta functions to both spread forces onto the background grid and interpolate velocities onto the immersed structure.  

\subsection{Regularized Delta Functions}\label{sec:reg_delta_fun}
\begin{figure}[t!]
    \centering
    \includegraphics[width=\textwidth]{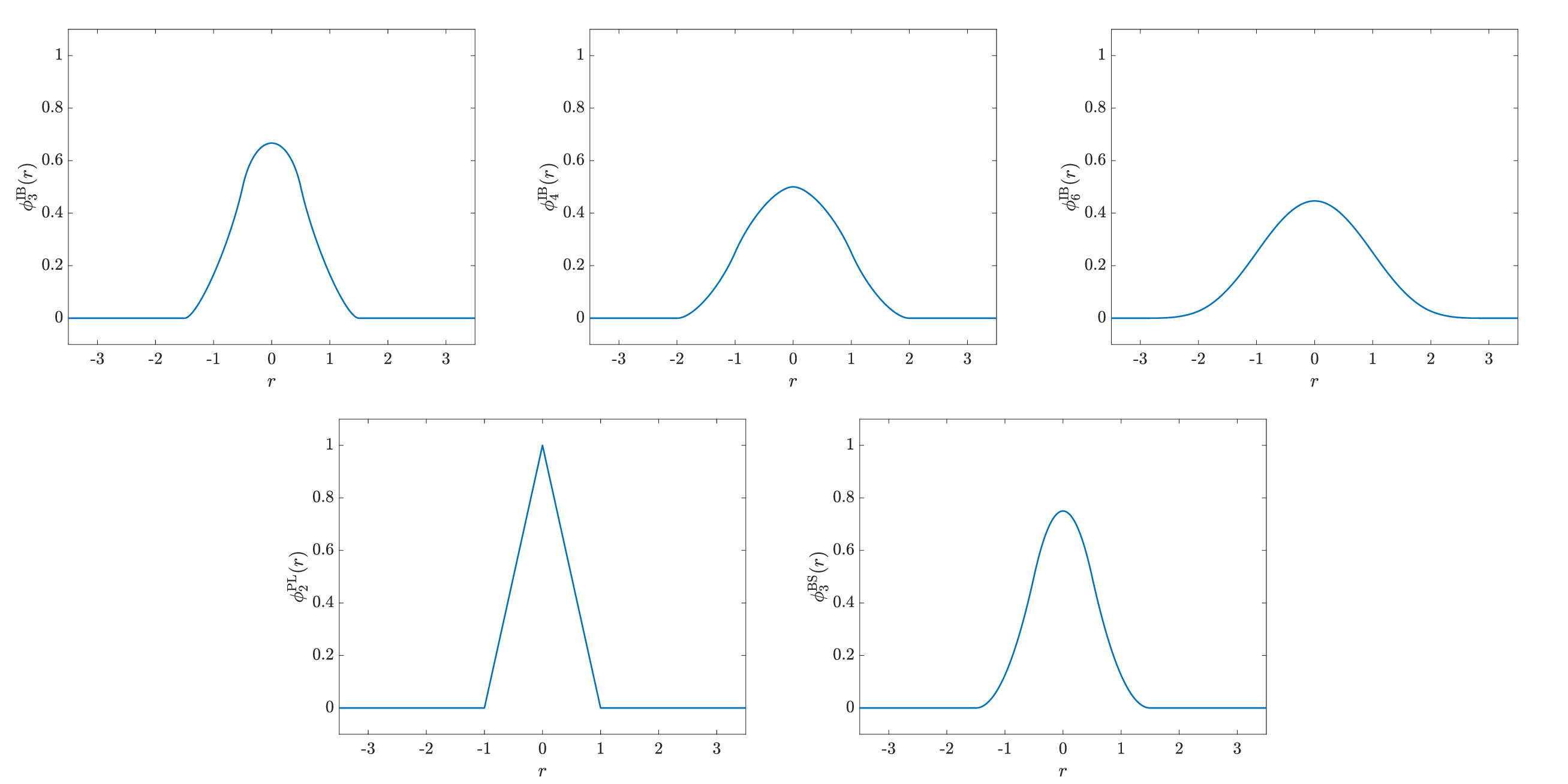}
    \caption{Illustrations of the kernel functions used to construct the regularized delta functions employed in this work. On the top from left to right are the three-point (IB3), four-point (IB4), and Gaussian-like six-point (IB6) IB kernel functions. Below from left to right are the piecewise linear (PL) and three-point B-spline (BS3) kernel functions.}
    \label{fig:kernel_functions}
\end{figure}

The computational tests in section \ref{numerical tests} explore the impact of the form of the regularized delta function on the accuracy of the numerical method. We consider five different constructions of $\delta_h$. Each construction is of a tensor-product form, so that
\begin{equation}
    \delta_h(\mathbf{x}) = \frac{1}{h^2}\phi\parens{\frac{x_1}{h}}\phi\parens{\frac{x_2}{h}},
\end{equation}
for different choices of basic one-dimensional kernel functions $\phi(r)$. The regularized delta functions we utilize are derived from two different families of one-dimensional kernel functions $\phi$. The first two kernel functions are the linear and quadratic members of the B-spline family. We refer to these kernel functions below as the piecewise-linear (PL) and three-point B-splines (BS3), respectively. The piecewise-linear B-spline kernel is the least regular of all the kernel functions considered in this study and is continuous but not differentiable. The three-point B-spline kernel is continuously differentiable\footnote{The discovery of the B-spline family of kernel functions is attributed to Schoenberg \cite{schoenberg1946}; however, even Schoenberg himself noted that Hermite, Peano, and Laplace were all well aware of B-splines\cite{schoenberg1973}.}. The other three kernel functions we consider belong to the so called IB family of kernel functions. The IB family of kernel functions were conceived by Peskin, who postulated the properties the IB family ought to satisfy to be computationally efficient, accurate, physical, and mathematically simple \cite{peskin2002}. In this work, we use the three-point, four-point, and Gaussian-like six-point IB kernel functions. The three-point IB (IB3) kernel was originally introduced in an adpative version of the IB method in which the Eulerian components were discretized on a staggered-grid \cite{roma1999}. The four-point IB (IB4) kernel is the most commonly used and satisfies all of Peskin's original postulates \cite{peskin2002}. The four-point IB and three-point IB kernels are both continously differentiable. The Gaussian-like six-point IB (IB6) kernel was formulated more recently \cite{bao2016} and is three times continuously differentiable, making it the smoothest kernel function that we use in this study. Figure \ref{fig:kernel_functions} displays each of the one dimensional kernel functions $\phi(r)$ we consider in this work. Further details regarding the properties of each of the regularized delta functions considered here can be found in Lee and Griffith's analysis of the IFED method\cite{lee2022}.

\subsection{Temporal Discretization}
We discretize in time using a modified trapezoidal rule for the momentum equation and imcompressibility constraint, the midpoint rule for the immersed boundary, and an explicit trapezoidal rule for the evolution of the conformation tensor. Specifically, to advance from time $t^n$ to time $t^{n+1} = t^n + \Delta t$, we first compute an approximation to the position of the immersed boundary and the conformation tensor at time $t^{n+1}$,
\begin{align}
    \frac{\widetilde{\coordMap}^{n+1} -\coordMap^n}{\Delta t}  &= \interp\left[\coordMap^{n}\right] \velocity^n,\\
    \frac{\widetilde{\conformTens}^{n+1}-\conformTens^n}{\Delta t} + \mathbb{N}_2^n &= \grad_h \velocity^n \conformTens^n + \conformTens^n \parens{\grad_h\velocity^n}\tran + \relaxationFunc\parens{\conformTens^n},
\end{align}
in which $\mathbb{N}_2^n$ is the approximation of $\velocity^n\cdot\grad\conformTens^n$ based on the second-order wave propagation algorithm described by Ketcheson et al.\cite{ketcheson2013} We then compute the position of the immersed boundary and Lagrangian force at the mid-time point $t^{n+\frac{1}{2}}$,
\begin{align}
    \coordMap^{n+\frac{1}{2}} = \frac{\widetilde{\coordMap}^{n+1} + \coordMap^n}{2}, \\
    \lagForce^{n+\frac{1}{2}} = \kappa\parens{\coordMap^0 - \frac{\widetilde{\coordMap}^{n+1} + \coordMap^n}{2}}.
\end{align}
We then obtain the new velocity and pressure,
\begin{align}
    \frac{\velocity^{n+1}-\velocity^n}{\Delta t} + \frac{3}{2} \mathbf{N}_1^n - \frac{1}{2}\mathbf{N}_1^{n-1} &= -\grad_h p^{n+\frac{1}{2}} + \fluidViscosity\grad^2_h \frac{\velocity^{n+1}+\velocity^n}{2} + \frac{\polymerViscosity}{\relaxationTime}\grad_h\cdot\parens{\frac{\widetilde{\conformTens}^{n+1}+\conformTens^n}{2}} + \spread\left[\coordMap^{n+\frac{1}{2}}\right]\lagForce^{n+\frac{1}{2}}, \\
    \grad_h\cdot\velocity^{n+1} &= 0,
\end{align}
in which $\mathbf{N}_1^n$ is an approximation of the convective term $\velocity^n\cdot\grad\velocity^n$ based on the xsPPM7 scheme\cite{griffith2009}. The resulting linear system is solved by a GMRES solver which applies the projection method as a preconditioner \cite{griffith2009}. Finally, we compute the conformation tensor and the immersed boundary position at time $t^{n+1}$ using an explicit trapezoidal rule and midpoint rule respectively,
\begin{align}
    \frac{\conformTens^{n+1}-\conformTens^n}{\Delta t} + \frac{1}{2}\parens{\mathbb{N}_2^n + \mathbb{N}_2^{n+1}} =& \frac{1}{2}\parens{\grad_h \velocity^n \conformTens^n + \conformTens^n \parens{\grad_h\velocity^n}\tran} + \frac{1}{2}\parens{\grad_h \velocity^{n+1} \widetilde{\conformTens}^{n+1} + \widetilde{\conformTens}^{n+1} \parens{\grad_h\velocity^{n+1}}\tran} \nonumber\\
    \quad & + \frac{1}{2}\parens{\relaxationFunc\parens{\conformTens^n} + \relaxationFunc\parens{\widetilde{\conformTens}^{n+1}}}, \\
    \coordMap^{n+1} =& \coordMap^n + \Delta t \interp\left[\coordMap^{n+\frac{1}{2}}\right]\velocity^{n+\frac{1}{2}},
\end{align}
in which $\velocity^{n+\frac{1}{2}} = \frac{1}{2}\parens{\velocity^n + \velocity^{n+1}}$ and $\mathbb{N}_2^{n+1}$ is an approximation of $\velocity^{n+1}\cdot\grad\widetilde{\conformTens}^{n+1}$. We note that each time step requires the solution of the Navier-Stokes equations, one force evaluation, and two velocity interpolation operations.

\subsection{Choice of the Spring Penalty Parameter $\kappa$}\label{penalty_spring_scaling}
In practice, the spring stiffness parameter $\kappa$ is determined to be the smallest value that keeps the Lagrangian markers from moving a Euclidean distance of more than $\frac{h}{2}$ from their initial position. When implementing convergence tests, we set 
\begin{equation}\label{eq:scaling_penalty}
\kappa = C\frac{\rho h}{\Delta t^2},
\end{equation}
in which $C$ is a dimensionless constant that is determined by finding the working $\kappa$ value for the coarsest discretization chosen. There are two reasons to scale $\kappa$ in this way: first, the scaling is consistent with the physical units of a linear spring; and second, it has been demonstrated in both by Lai \cite{lai1998} and by Hua and Peskin \cite{hua2022} that this scaling of $\kappa$ maintains the numerical stability of the IB system of equations in the context of the Navier-Stokes equations. In the viscoelastic fluids simulations carried out here, we also observe that this scaling $\kappa$ maintains the stability of our numerical scheme when making refinements to the grid and time step size.

\subsection{Definition of Discrete $L^p$ norms}
To empirically analyze the convergence properties of the IB method, we make use of discrete $L^p$ norms. The discrete $L^p$ norms used herein are the same as the ones defined by Griffith et al.\cite{griffith2007}.

\subsection{Software Implementation}
The solvers are implemented in IBAMR \cite{ibamr}, which is an MPI-parallelized implementation of the immersed boundary method. Support for structured adaptive mesh refinement is provided by SAMRAI \cite{hornung2002,hornung2006}, and the linear solvers are implemented using PETSc \cite{petsc_users_manual,petsc_web_page,balay1997}. 
\section{Numerical Tests}\label{numerical tests}
We use the IB method to model the flow of a viscoelastic fluid through or around a stationary structure. The rigidity of the structure is weakly imposed using the penalty spring forcing approach as described in section \ref{sec:model_description}. Initially, the viscoelastic fluid is assumed to be at rest and the dynamics of the flow are driven by the specification of inflow and outflow velocity boundary conditions. For each test, the simulation is allowed to proceed until steady-state conditions are reached. We found empirically that steady-state conditions are achieved when the duration of the simulation exceeds twenty times the size of the relaxation number $\relaxationTime$ in the case of the Oldroyd-B fluid model, or twenty times the size of the reptation time $\reptationTime$ in the case of the Rolie-Poly fluid model. Additionally, for each numerical test, we perform a grid convergence study and estimate rates of convergence either by comparing our computed solution to an analytic steady-state solution when available, or by utilizing Richardson extrapolation. When refinements to the grid are made, the timestep size and penalty spring parameter $\kappa$ are scaled consistently to ensure the stability of our numerical scheme. The timestep size is scaled as some constant multiplied by finest Cartesian grid increment. The penalty spring parameter $\kappa$ is set according to equation \eqref{eq:scaling_penalty}, and we report only dimensionless constant $C$ in each of the numerical tests below. 

\subsection{Oldroyd-B Flow Through an Inclined Channel}
We begin our series of numerical tests by analyzing Oldroyd-B flow through an inclined channel. The channel was chosen to be inclined rather than completely vertical or horizontal to avoid having a completely grid aligned discretization of the channel walls. To assess the accuracy of our numerical simulations, we simulate startup Oldroyd-B channel flow using the steady-state solution for Oldroyd-B channel flow as the inflow boundary condition and allow the simulation to proceed until a given simulation until steady state is reached. The absolute errors associated with our steady-state solution is then computed in the $L^1$, $L^2$, and $L^{\infty}$ grid norms. \par
\begin{figure}[t!]
\centering
\includegraphics[scale = 0.5]{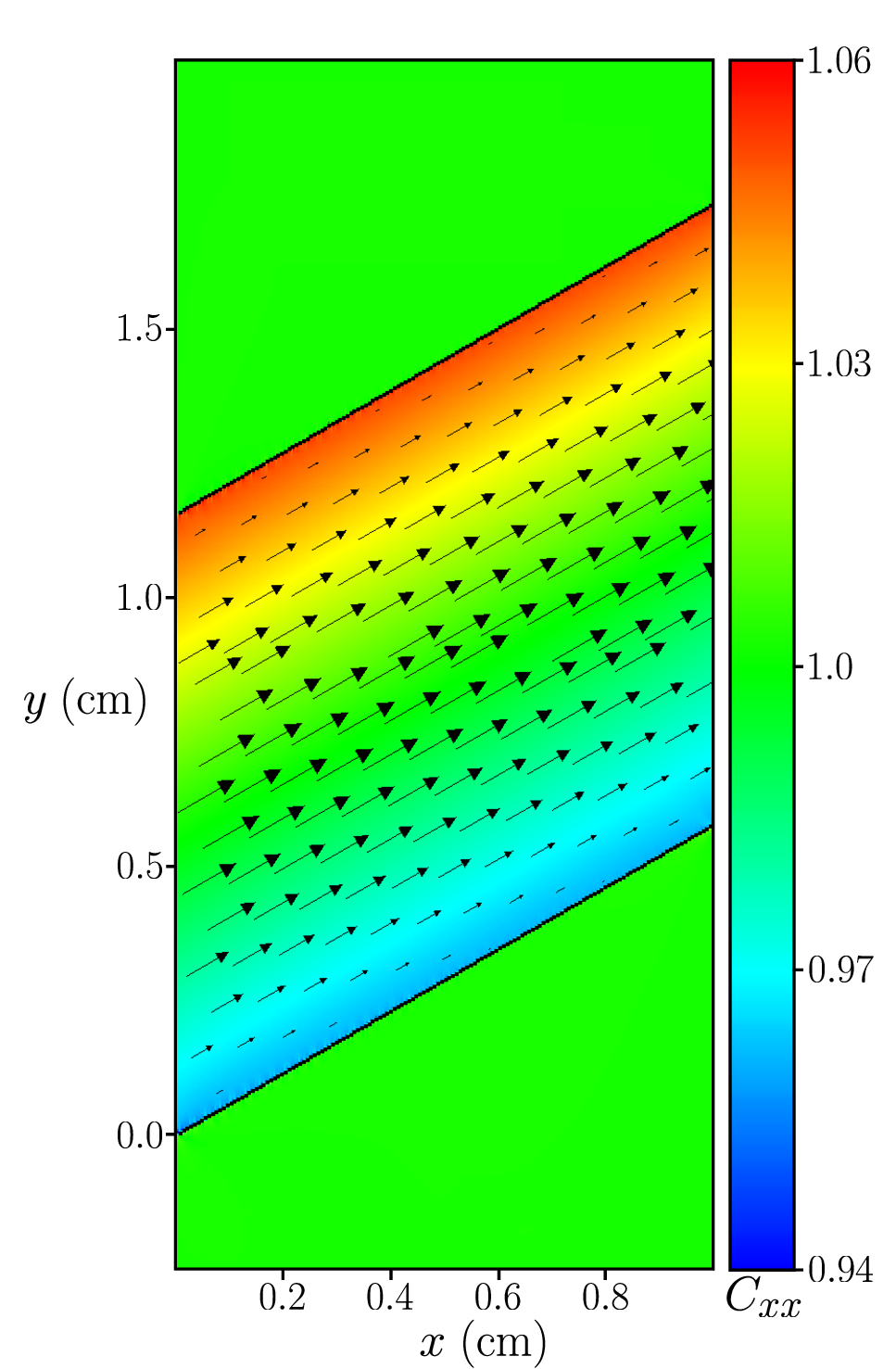}
\caption{Visualization of the computed steady state solution to Oldroyd-B flow through an inclined channel using piecewise linear kernel with $M_{\text{fac}} = 2.0$, $\mu_{\text{s}} = \mu_{\text{p}} = \SI{0.05}{\pascal\cdot\s}$, $\theta = \frac{\pi}{6}$, and $\lambda = \SI{0.1}{\s}$ on the finest uniform grid ($h = \frac{H}{256}$). The velocity vector field (in black) is illustrated alongside the (1,1) component of the dimensionless conformation tensor $\conformTens$. The units of length are in centimeters. The black dots outlining the channel represent the steady state locations of the Lagrangian markers.}
\label{fig:Slanted_Channel_flow}
\end{figure}

\begin{figure}[t!]
\centering
\begin{subfigure}
\centering 
\includegraphics[width=.45\linewidth]{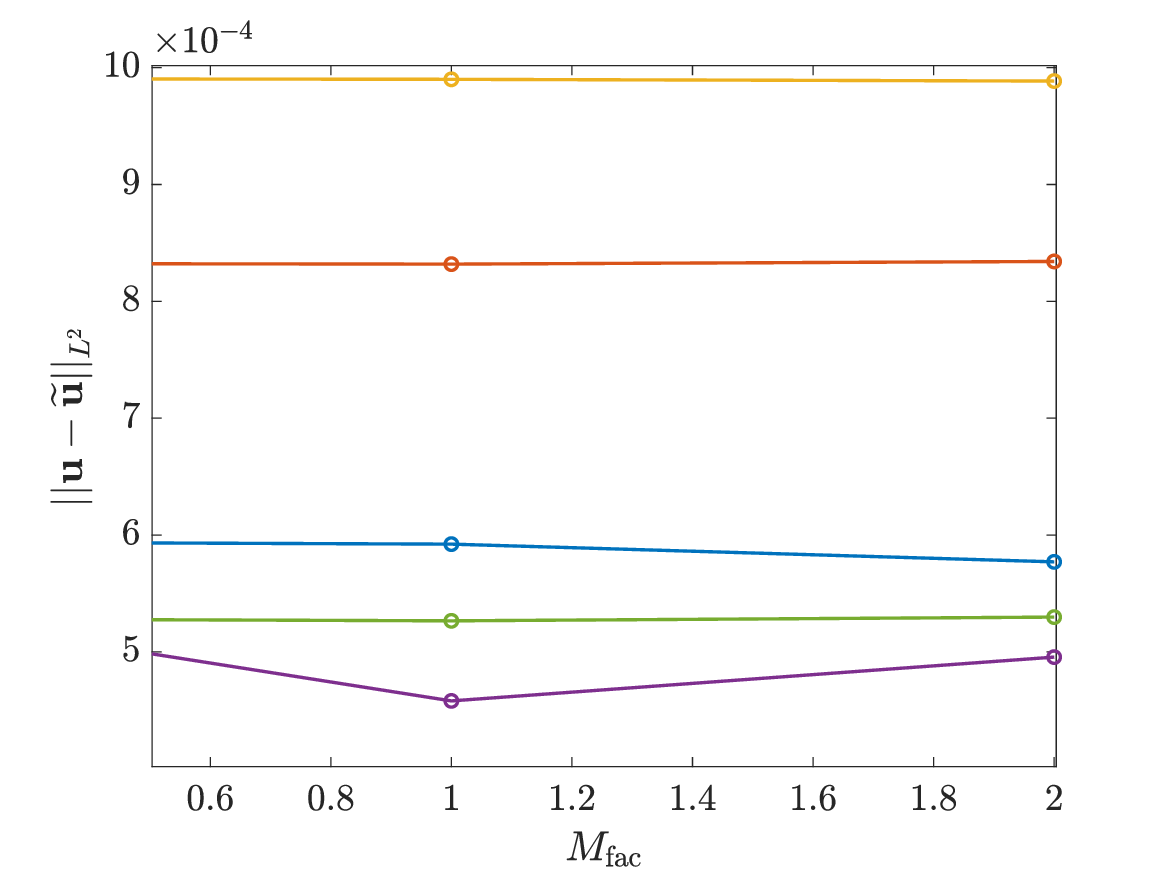}
\end{subfigure}
\begin{subfigure}
\centering 
\includegraphics[width=.45\linewidth]{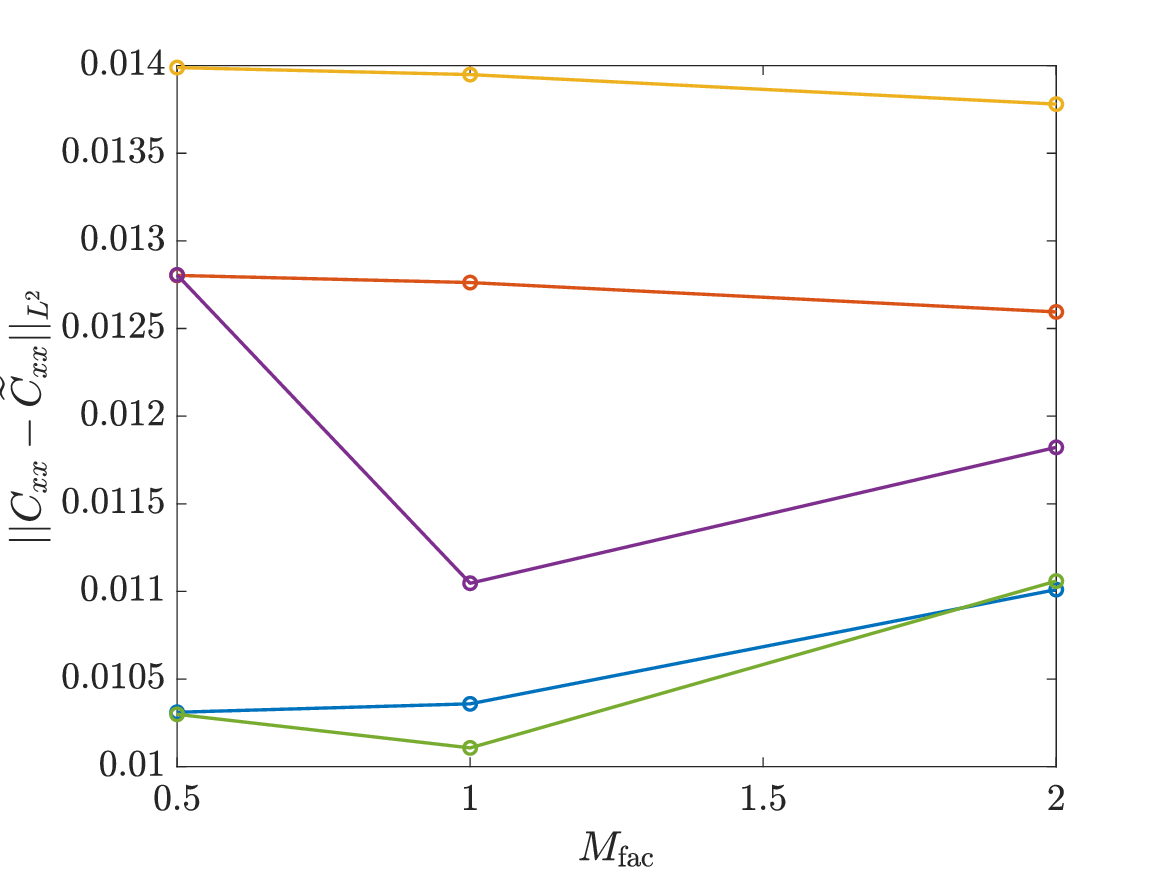}
\end{subfigure}
\\
\begin{subfigure}
    \centering
    \includegraphics[scale = 0.45]{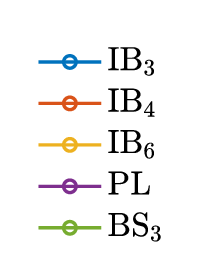}
\end{subfigure}
\\
\begin{subfigure}
    \centering
    \includegraphics[width=0.45\linewidth]{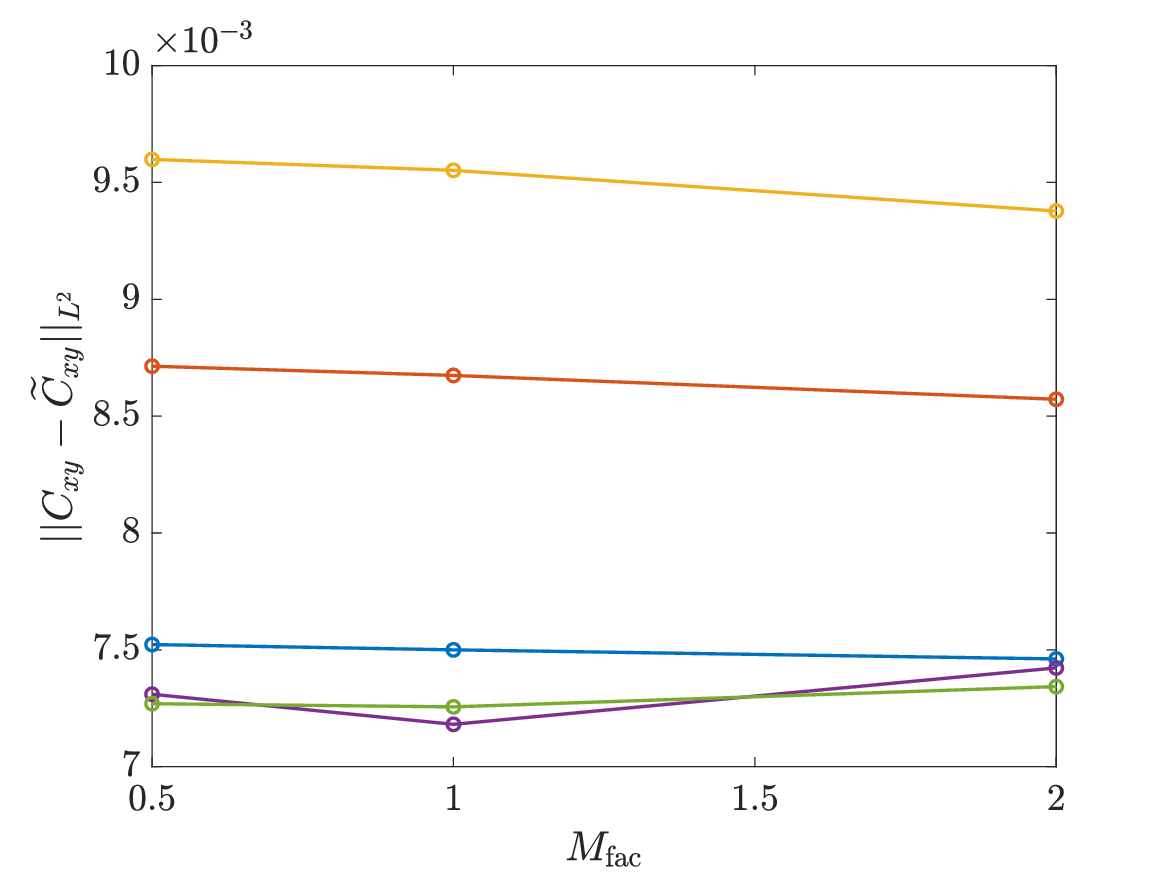}
\end{subfigure}
\begin{subfigure}
    \centering
    \includegraphics[width=0.45\linewidth]{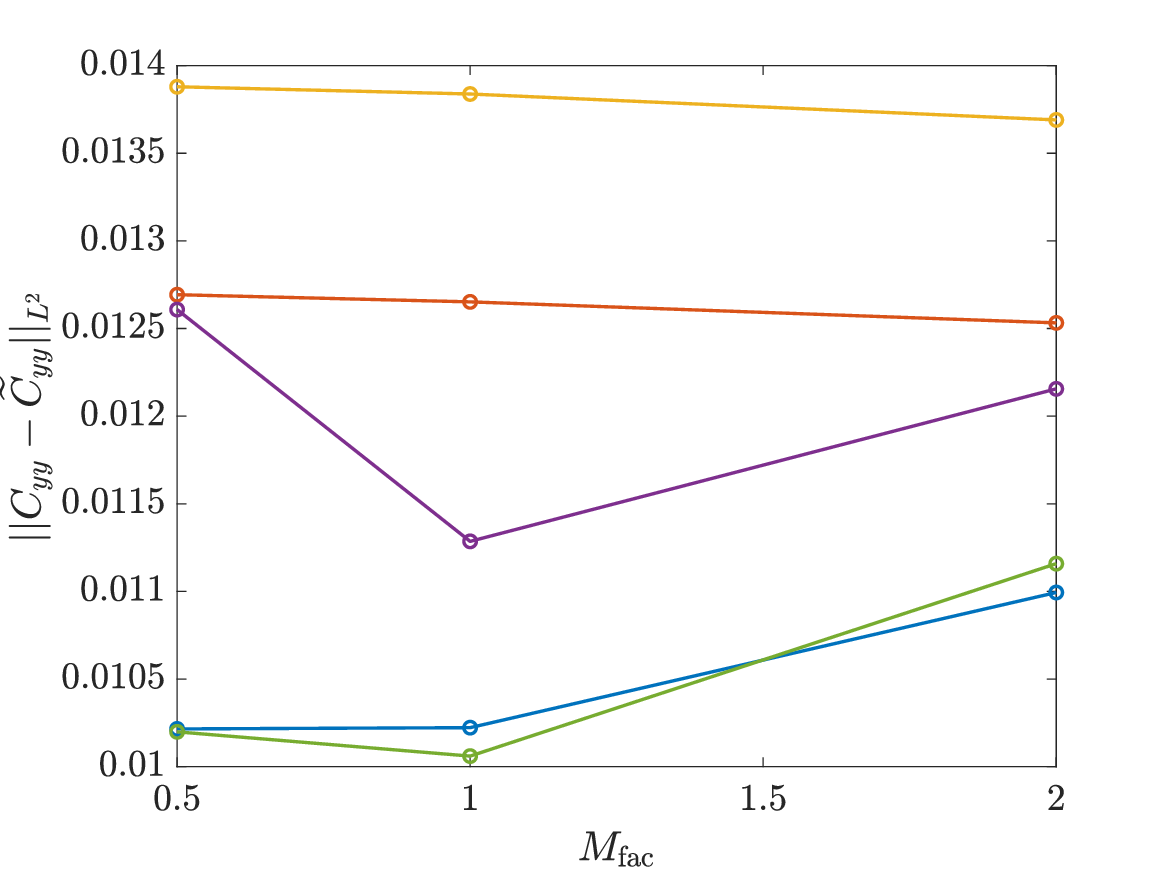}
\end{subfigure}
\caption{Plots of $L^2$ grid errors at the intermediate discretization of $h = \frac{H}{128}$ for working $M_{\text{fac}}$ values of $\frac{1}{2},1,$ and $2$. For each $M_{\text{fac}}$ value tested, the computed steady state values of the velocity $\tilde{\mathbf{u}}$ and the conformation tensor $\tilde{\conformTens}$ were computed and compared against the against their respective analytic solutions for Oldroyd-B flow through a channel. We observe that the $L^2$ error is independent of the $M_{\text{fac}}$ values tested. The error dependence on the $M_{\text{fac}}$ values tested in the $L^{\infty}$ and $L^1$ norms are analgous.}
\label{Error_Analysis_Slanted_Channel}
\end{figure}

 
Assuming a constant pressure gradient of unit magnitude, the analytic steady-state solution to Oldroyd-B flow through the inclined channel is 
\begin{align}
 &   u(x,y) = -\frac{\cos\theta}{2\left(\mu_{\text{s}} + \mu_{\text{p}}\right)}\left(y\cos\theta - x\sin\theta\right)\left(y\cos\theta - x\sin\theta - H\right),\\ 
 &   v(x,y) = -\frac{\sin\theta}{2(\mu_{\text{s}} + \mu_{\text{p}})}\left(y\cos\theta - x\sin\theta\right)\left(y\cos\theta - x\sin\theta - H\right), \\
 &   C_{xx}(x,y) = \cos^2\theta\tilde{C}_{xx}(x,y) + \sin^2\theta\tilde{C}_{yy}(x,y) - 2\cos\theta\sin\theta\tilde{C}_{xy}(x,y), \\
 & C_{xy}(x,y) = \tilde{C}_{xy}(x,y)\left(\cos^2\theta - \sin^2\theta\right) + \cos\theta\sin\theta\left(\tilde{C}_{xx}(x,y) - \tilde{C}_{yy}(x,y)\right) ,\\
 & C_{yy}(x,y) = \cos^2\theta\tilde{C}_{yy}(x,y) + \sin^2\theta\tilde{C}_{xx}(x,y) + 2\cos\theta\sin\theta\tilde{C}_{xy}(x,y),
\end{align}
in which $H$ is the height of the horizontal channel, $\theta$ is the angle of rotation of the horizontal channel about the origin, and $\tilde{\conformTens}$ is the conformation tensor associated with the analytic steady state solution for the horizontal channel,
\begin{align}
& \tilde{C}_{xx}(\tilde{x},\tilde{y}) = 1 + 2\tilde{C}_{xy}(\tilde{y})^2, \\
& \tilde{C}_{xy}(\tilde{x},\tilde{y}) = \lambda\frac{\partial u}{\partial \tilde{y}} = -\frac{\lambda}{\mu_{\text{s}}+\mu_{\text{p}}}\left(\tilde{y} - \frac{H}{2}\right), \\
& \tilde{C}_{yy} = 1.
\end{align}
We set $\theta = \frac{\pi}{6}$, $H = \SI{1.0}{\centi\meter}$, $\mu_{\text{s}} = \mu_{\text{p}} = \SI{0.05}{\pascal\cdot\s}$, $\rho = 1.0$ $\unit[per-mode=symbol]{\gram\per\cm^3}$, and $\lambda = \SI{0.1}{\s}$. All simulations are run until a final time of $T_{\text{final}} = 20\lambda = \SI{2.0}{\s}$ to ensure that steady state is approximately reached. For each choice of regularized delta function and $M_{\text{fac}}$ value, we perform a total of four simulations on increasingly fine uniform grids $h = \frac{H}{32}, \frac{H}{64}, \frac{H}{128},$ and $\frac{H}{256}$. The time step size for each discretization is chosen to be $\Delta t = 0.2h\,\si{\s}$, and the stiffness parameter $\kappa$ is set by equation \eqref{eq:scaling_penalty} with a dimensionless constant of $C=7.5\times10^{-2}$. Figure~\ref{fig:Slanted_Channel_flow} shows a representative solution.
 Figure~\ref{Error_Analysis_Slanted_Channel} analyzes the impact of $M_{\text{fac}}$ on the accuracy of the computed solution. The accuracy of the computed solution appears to be independent of the working $M_{\text{fac}}$ value chosen. We observe that the use of regularized delta functions with narrower supports provide more accurate solutions at comparable grid resolutions than regularized delta functions of broader support.  At first, this finding may seem somewhat surprising since all the regularized delta functions considered satisfy the same discrete moment conditions, which should make them all approximately equally accurate interpolants of \textit{smooth} functions.  However, the velocity we are approximating with the IB method is not a smooth function, because the true velocity has a discontinuous jump in its normal derivative across the immersed boundary \cite{lai2001}. The discrepancies in errors across different regularized delta functions can be explained, at least heuristically, by considering the interpolation error associated with a single-variable function whose normal derivative has a jump discontinuity at the interpolation point $x_0$ of interest. For such a function, the leading order error term is precisely the mean value of the regularized delta function taken over the half-line $[x_0,\infty)$ or $(\infty,x_0]$ (the regularized delta function is assumed to be even) multiplied by the value of the jump in the function's derivative at the interpolation point $x_0$. This finding was applied originally by Beyer and Leveque \cite{beyer1992} to derive a second-order accurate one-dimensional version of the IB method. Consequently, non-negative regularized delta functions with narrower support (and hence smaller mean values on the half-line) will provide more accurate interpolations of a velocity that is $C^0$ but not $C^1$ the point of interest. In the context of this study, we note that the half space mean value of the six point IB kernel is approximately twice that of the piecewise linear kernel, and, as Figure \ref{Error_Analysis_Slanted_Channel} demonstrates, the error in the computed velocity using the six-point IB regularized delta function is approximately twice that of the piecewise-linear regularized delta function.  \par
\begin{figure}[t!]
    \centering
    \includegraphics[width=1.0\linewidth]{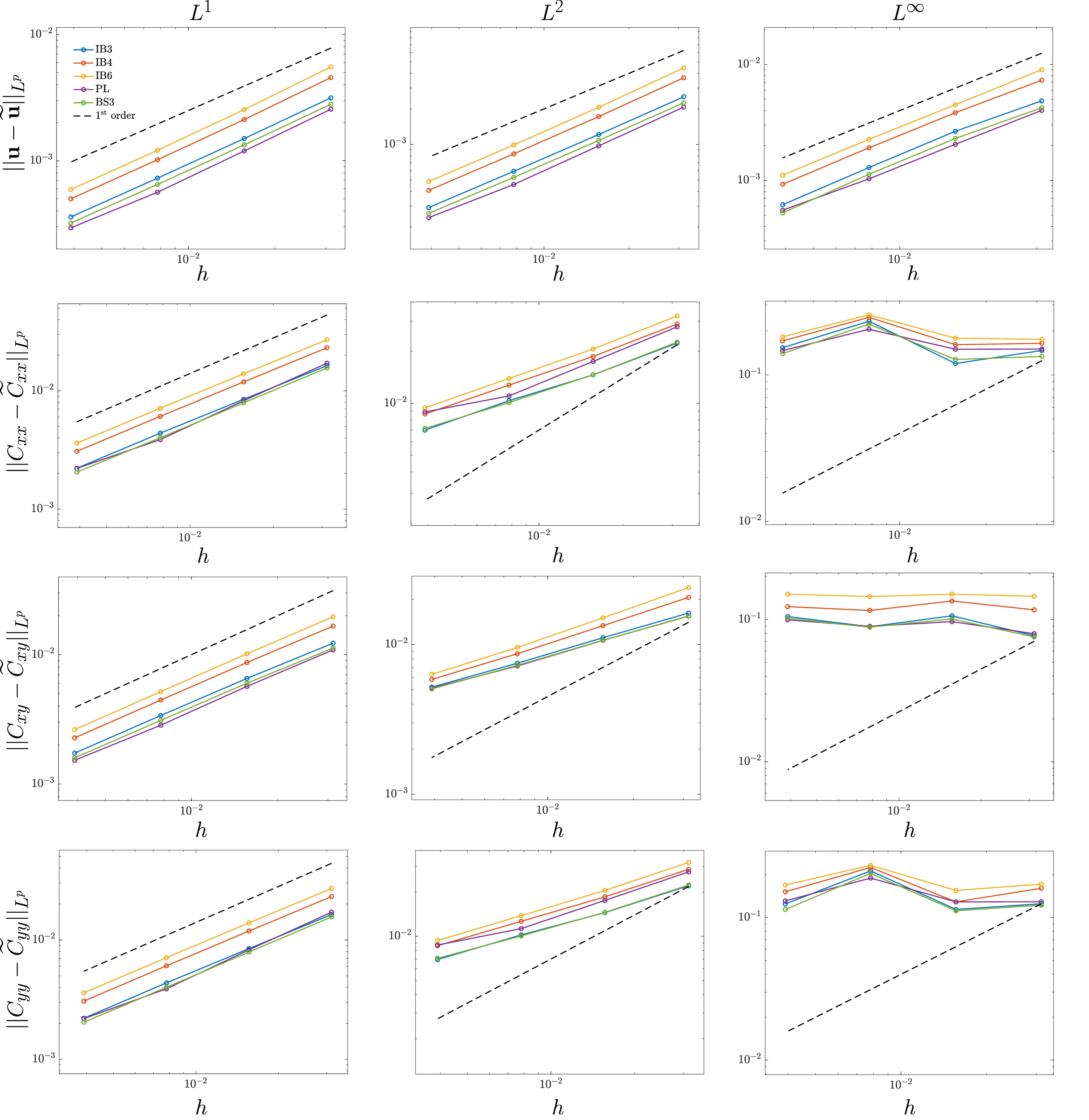}
    \caption{Log-log plots analyzing the convergence of the computed steady-state velocity $\mathbf{u}$ and conformation tensor $\conformTens$ to the analytic steady-state velocity $\widetilde{\mathbf{u}}$ and conformation tensor $\widetilde{\conformTens}$ for Oldroyd-B flow through an inclined channel. For each simulation, a $M_{\text{fac}}$ value of 1 was chosen. We observe first-order convergence in all grid norms for the components of the velocity and approximately first-order convergence in the $L^1$ norm, a half-order of convergence in the $L^2$ norm, and the failure of convergence in the $L^{\infty}$ norm for each component of the conformation tensor. Convergence results for the other $M_{\text{fac}}$ values tested were similar.}
    \label{Convergence_M_fac_2_Slanted_Channel}
\end{figure}

\begin{figure}[t!]
    \centering
    \includegraphics[scale=0.5]{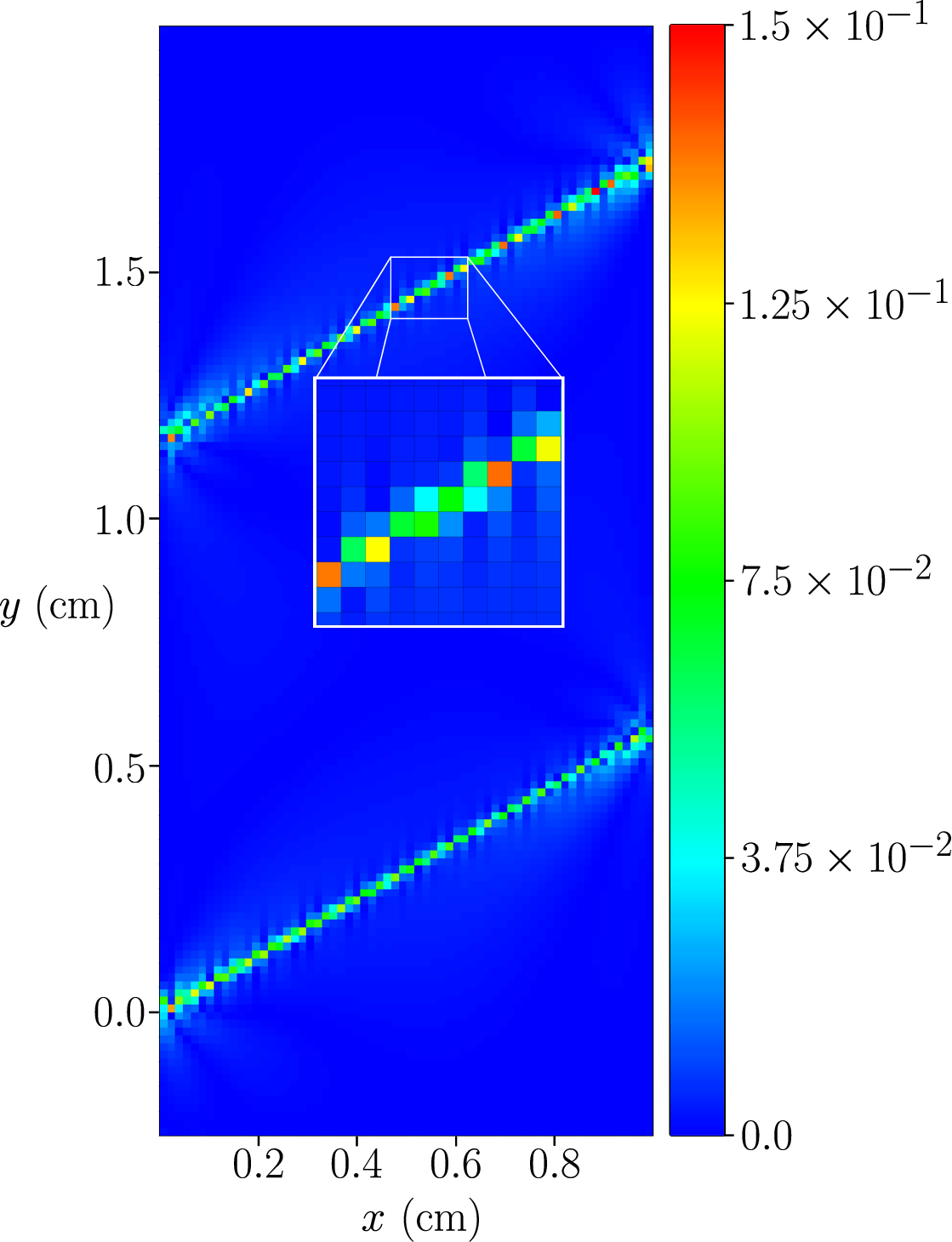}
    \caption{The absolute error in the $C_{xx}$ component of the conformation tensor for the Oldroyd-B flow through a slanted channel using the three-point B-spline regularized delta function at the intermediate discretization of $h = \frac{H}{64}$ with $M_{\text{fac}} = 2$. The largest errors are contained in a small region, about two to four Cartesian grid increments wide, where the conformation tensor fails to converge pointwise.}
    \label{fig:Error_Cxx_Pseudo_color}
\end{figure}

 
Figure \ref{Convergence_M_fac_2_Slanted_Channel} analyzes the global convergence properties of the IB method applied to the Oldroyd-B flow through an inclined channel for $M_{\text{fac}} = 2$. In general, we find that the velocity converges at a first-order rate for each of the grid norms used while the components of the conformation tensor converge at about a first-order rate in $L^1$, a half-order rate in $L^2$, and fails to converge in $L^{\infty}$. The convergence rates for the other $M_{\text{fac}}$ values tested are consistent with these results. The conformation tensor's lack of pointwise convergence is a consequence of the lack of pointwise convergence of the velocity gradients in the IB method. Figure \ref{fig:Error_Cxx_Pseudo_color} complements Figure \ref{Convergence_M_fac_2_Slanted_Channel} by illustrating the absolute error in computing the $C_{xx}$ component of the conformation tensor using the IB method. Figure \ref{fig:Error_Cxx_Pseudo_color} demonstrates that the largest errors are localized to a region about two to four Cartesian grid increments wide surrounding the immersed boundary. It is in this region where the velocity gradient computed using the IB method fail to converge pointwise. Since the evolution of the conformation tensor is directly related to the velocity gradient, the components of the conformation tensor fail to converge pointwise in this region as well. We note that, although regularized delta functions with wider supports generally generate larger errors in the computation of the conformation tensor, the width of the region where pointwise convergence fails appears to independent of the regularized delta function employed. Although the IB method is unable to obtain pointwise values for the conformation tensor near the immersed boundary, we show in section \ref{sec:OlroydBcylinder} that the IB method is still able to capture convergent \textit{net forces} on structures immersed in viscoelastic flow.


\subsection{Oldroyd-B Flow Past a Cylinder}\label{sec:OlroydBcylinder}
Next, we utilize the IB method to study the problem of Oldroyd-B flow past a confined and stationary cylinder. For this problem, we specify the cylinder to have a radius of $R=\SI{1}{\mm}$, and we initialize its location to the center of a channel $\Omega=[-L/2,L/2]\times[-H/2,H/2]$ with $L=32R$ and $H = 4R$. To be consistent with previous literature studies, we work with a dimensionless system of equations in which the spatial coordinates are rescaled by the radius $R$, the velocity is rescaled by a characteristic velocity $U$, and time is rescaled by the ratio $\frac{R}{U}$. The pressure is rescaled by $\frac{\mu}{R}$, in which $\mu = \mu_s+\mu_p$, and the Eulerian force density $\eulForce$ is rescaled by $\frac{U\mu_s}{R^2}$ which, leads to the Lagrangian force density $\lagForce$ being rescaled by $\frac{\mu_s U}{R}$. After an application of the chain rule and some algebraic manipulations, the resulting dimensionless system of equations is
\begin{equation}
\Re\parens{\frac{\partial\velocity\parens{\xx,t}}{\partial t} + \velocity\parens{\xx,t}\cdot\grad\velocity\parens{\xx,t}} = -\grad p +  \beta\Delta\velocity + \frac{1-\beta}{\Wi}\grad\cdot\conformTens +  \beta\eulForce\parens{\xx,t},
\end{equation}
in which $\beta = \frac{\fluidViscosity}{\viscosity}$. In each of our simulations the parameter $\beta$ is set to a value of $\beta = 0.59$, which aligns with previous benchmark studies. We employ no-slip conditions for the velocity $\parens{\velocity = \mathbf{0}}$ along with homogeneous Neumann boundary conditions for the conformation tensor on both the top and bottom walls of the channel. At the inflow boundary, the steady-state Oldroyd-B pipeflow solution is implemented, corresponding to the parabolic velocity flow profile $\velocity = \frac{3}{2}\parens{1 - \frac{y^2}{2}}$. At the outflow boundary, we set the pressure equal to zero and stipulate homogenous Neumann boundary conditions for the conformation tensor. 

\begin{figure}[t!]
    \centering
    \includegraphics[width=\textwidth]{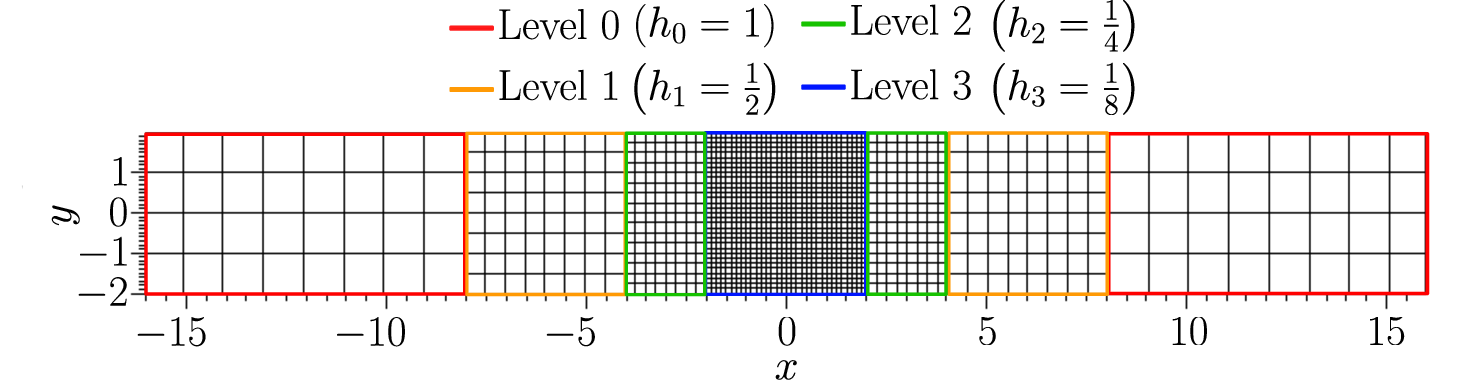}
    \caption{A representative illustration of the statically refined grid used in the Oldroyd-B flow past a cylinder.}
    \label{fig:static_grid_cylinder}
\end{figure}

Due to the relatively long length of the channel, we simulate the resulting equations of motion on a static, locally refined Cartesian grid. The locally refined grid consists of four different levels of spatial refinement $h_0,h_1,h_2,$ and $h_3$ which differ from one another by a refinement ratio of $r = 2$. A coarsened version of the locally refined grid we utilize is displayed in Figure \ref{fig:static_grid_cylinder}.
The second-order accurate spatial discretization of the Eulerian variables $\velocity$, $p$, and $\conformTens$ described in section \ref{sec:space_disc} is adapted to the locally refined Cartesian grid by following the discretization procedures outlined by Griffith\cite{griffith_heart2012}. In the literature, Oldroyd-B flow past a confined stationary cylinder is typically studied in the zero Reynolds number limit which couples the extra-stress or conformation tensor to the steady Stokes equations rather than the full Navier-Stokes equations \cite{stein2019,dou1999,alves2001,fan2005}. However, our discretization near the coarse-fine interfaces of the locally refined Cartesian grid limits the solvers used herein to non-zero Reynolds numbers. Consequently, we perform the following simulations at the Reynolds number of $\text{Re}=1.0\times 10^{-5}$ to make comparisons to literature values.

For each regularized delta function and $M_{\text{fac}}$ value, we conduct simulations on a series of increasingly refined grids with the finest Cartesian grid increment $h_3$ set to $\frac{1}{16},\frac{1}{32}$, and $\frac{1}{64}$. The timestep size for each grid is $\Delta t = \left(7.5\times10^{-4}\right)h_3$. The spring penalty parameter $\kappa$ is adjusted according to equation \eqref{eq:scaling_penalty} using a dimensionless constant of $C = 2.95$, and is further rescaled by $\frac{U\mu_s}{R^2}$ so that it is consistent with the dimensionless formulation presented. Each simulation is run until $T_{\text{final}} = 20\Wi$ so that a steady-state solution is achieved.\par 

To assess the accuracy of our numerical simulations, we compute convergence rates of the velocity, conformation tensor, and pressure fields and also compute the steady-state drag coefficient. We focus on the well studied low Weissenberg number case $\text{Wi} = 0.1$, for which the consensus literature value is approximately $C_{\text{d}} = 130.364$ \cite{stein2019}. To compute the steady-state drag coefficient, we follow the method employed in by Stein et al.\cite{stein2019} and compute the steady-state drag coefficient by a discretization of
\begin{equation}
    C_{\text{d}} = -\beta\int_{\Gamma}\mathbf{F}(\mathbf{X})\cdot \hat{\mathbf{x}}\,\text{d}A,
\end{equation}
  in which $\hat{\mathbf{x}}$ is the unit vector pointing in the $x$-direction. We compute the drag in this a way to avoid computing a surface integral defined in terms of the stress because, as mentioned above, the conformation tensor and the extra stress tensor are not expected converge along immersed boundaries when using the IB method. Table \ref{tab:Drag_Errors} reports the relative errors in the computed drag coefficients for each kernel and $M_{\text{fac}}$ value tested. In particular, we observe that for each of the regularized delta functions tested, the computed drag coefficient appears to converge at a first-order rate, indicating that the IB method is capable of computing accurate net forces on immersed boundaries as long as the grid resolution is fine enough. In addition, similar to Oldroyd-B flow through the inclined channel above, we observe that for the finest resolution cases, across all $M_{\text{fac}}$ values tested, using regularized delta functions of narrower support tends to lead to more accurate computations of drag coefficient. However, we also observe that the accuracy of the computed drag coefficient appears much more sensitive to the $M_{\text{fac}}$ value for regularized delta functions of narrower support. Preliminary results indicate that the sensitivity in the accuracy of the drag coefficient associated with the $M_{\text{fac}}$ value chosen is not primarily influenced by the presence of a viscoelastic fluid. Rather, it appears to be inherent to the IB method itself.
  
  %
  
\begin{figure}[t!]
    \centering
    \includegraphics[width=0.95\linewidth]{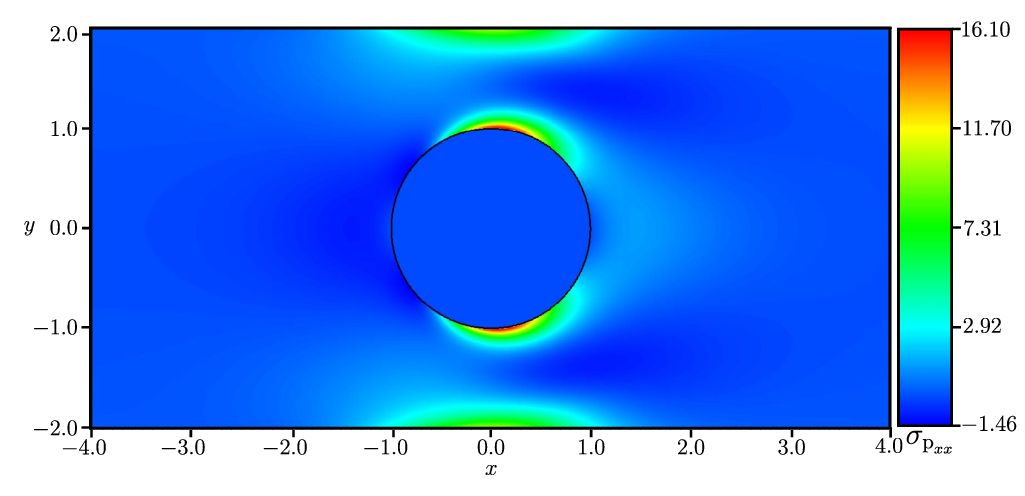}
    \caption{The computed steady-state solution of the $(1,1)$-component of the polymeric stress tensor $\sigma_{\text{p}_{xx}}$ for Oldroyd-B flow past at a Weissenberg number of $\text{Wi}=0.1$ using the same parameters as described in \cite{stein2019,dou1999,alves2001,fan2005}. IB markers outlining the boundary of the cylinder are displayed in black. Note that here we show only the section of the computational domain near the cylinder. The total length of the computational domain is $L = 32$.}
    \label{fig:stress_11_flow_past_cylinder}
\end{figure}

  \begin{table}[h]
  \caption{Relative Errors in the computed drag coefficient $C_{\text{d}}$ for Oldroyd-B flow past a stationary cylinder for each of the regularized delta functions and $M_{\text{fac}}$ values tested.}
    \centering
    \begin{tabular}{l c c c c c }
    \toprule
     $M_{\text{fac}} = \frac{1}{2}$ & BS3 & IB3 & PL & IB4 & IB6 \\\midrule
      $h_{3} = \dfrac{1}{16}$ & $6.15\%$ & $9.55\%$ & $12.92\%$ & $10.50\%$ & $11.16\%$ \\[1.5ex]
      $h_{3}  = \dfrac{1}{32}$ & $2.78\%$ & $4.11\%$ & $5.39\%$ & $4.73\%$ & $5.02\%$ \\[1.5ex]
      $h_{3} = \dfrac{1}{64}$ & $1.39\%$ & $2.04\%$ & $2.71\%$ & $2.32\%$ & $2.38\%$ \\[1.5ex]\midrule
      $M_{\text{fac}} = 1$ & ~ & ~ & ~ & ~\\\midrule
      $h_{3} = \dfrac{1}{16}$ & $5.45\%$ & $6.20\%$ & $4.77\%$ & $9.51\%$ & $11.16\%$ \\[1.5ex]
      $h_{3} = \dfrac{1}{32}$ & $2.61\%$ & $2.92\%$ & $2.54\%$ & $4.37\%$ & $5.02\%$ \\[1.5ex]
      $h_{3} = \dfrac{1}{64}$ & $1.24\%$ & $1.42\%$ & $1.19\%$ & $2.07\%$ & $2.38\%$ \\[1.5ex]\midrule
      $M_{\text{fac}} = 2$ & ~ & ~ & ~ & ~ \\\midrule
      $h_{3} = \dfrac{1}{16}$ & $4.57\%$ & $5.86\%$ & $1.57\%$ & $9.06\%$ & $11.16\%$ \\[1.5ex]
      $h_{3} = \dfrac{1}{32}$ & $2.12\%$ & $2.66\%$ & $1.03\%$ & $4.16\%$ & $5.02\%$ \\[1.5ex]
      $h_{3} = \dfrac{1}{64}$ & $1.01\%$ & $1.28\%$ & $0.46\%$ & $1.99\%$ & $2.38\%$ \\[1.5ex]
      \bottomrule

    \end{tabular}
   
    \label{tab:Drag_Errors}
\end{table}
  \begin{table}[h]
  \caption{Empirical global convergence rates of the velocity and components of the polymeric stress tensor $\polymerStress$ for Oldroyd-B flow past a cylinder using the three-point B-spline kernel and an $M_{\text{fac}}$ value of $\frac{1}{2}$. Empirical rates of convergence for other kernel functions using different $M_{\text{fac}}$ values are similar. Empirical rates of convergence were estimated using Richardson extrapolation.}
    \centering
    \begin{tabular}{c c c c }
    \toprule
     ~ & $L^1$ rate & $L^2$ rate & $L^{\infty}$ rate \\\midrule
     $\mathbf{u}$ & 1.693 & 1.470 & 1.020 \\[1.5ex]
     $\sigma_{\text{p}_{xx}}$  & 1.173 & 0.537 & -0.093\\[1.5ex] 
     $\sigma_{\text{p}_{xy}}$  & 1.341 & 0.632 & 0.130\\[1.5ex] 
     $\sigma_{\text{p}_{yy}}$  & 1.254 & 0.872 & 0.017 \\[1.5ex]
     $p$          & 1.275 & 1.204 & -0.541\\[1.5ex]
      \bottomrule
    \end{tabular}
    \label{convergence_rates_flow_past_cylinder}
\end{table}
To complement our analysis of Oldroyd-B flow past a cylinder, Table \ref{convergence_rates_flow_past_cylinder} shows the empirical global convergence rates of the velocity, stress tensor, and pressure computed using the three-point B-spline kernel with an $M_{\text{fac}}$ of $\frac{1}{2}$. Empirical rates for other regularized delta functions and choices of $M_{\text{fac}}$ are similar.  Below Table \ref{convergence_rates_flow_past_cylinder}, we present the steady-state solution of the $(1,1)$-component of the polymeric stress tensor in Figure \ref{fig:stress_11_flow_past_cylinder}. For a Weissenberg number of $\text{Wi}=0.1$, spectral discretizations of Oldroyd-B flow past a cylinder report the maximum value of the $(1,1)$ component of the dimensionless polymeric stress, $\sigma_{\text{p}_{xx}}$, to be between $17$ and $19$ \cite{stein2019,claus2013}. The piecewise-linear regularized delta function implemented with a $M_{\text{fac}}$ value of $\frac{1}{2}$ was the only instance which yielded an overestimate of the $(1,1)$-component of the polymeric stress, yielding a value of $||\sigma_{\text{p}_{xx}}||_{\infty} = 21.71$. Additionally, we find that maximum value of the $(1,1)$-component of the polymeric stress obtained using smoother regularized delta functions was less sensitive to the choice of $M_{\text{fac}}$. The least sensitive regularized delta function is the six-point, Gaussian-like regularized delta function, which yielded a consistent value of $||\sigma_{\text{p}_{xx}}||_{\infty} = 15.49$ for each of the $M_{\text{fac}}$ values tested. The most sensitive regularized delta function was the $C^0$ piecewise linear kernel function which, generated $||\sigma_{\text{p}_{xx}}||_\infty = 16.82,17.06,$ and $21.71$ for $M_{\text{fac}}$ values of $2,1,$ and $\frac{1}{2}$, respectively.  \par 

\FloatBarrier

\subsection{Rolie-Poly Flow Through a Cross Slot and a Contraction-Expansion Slit}
In this section, we utilize the IB method to model the flow of monodisperse polystyrene DOW1568 through two experimental flow geometries:a contraction-expansion slit and a cross-slot. The geometry of the cross slot and contraction expansion slit are shown in Figures \ref{fig:Contraction_Slit_geom} and \ref{fig:Cross_Slot_Geom}, respectively. We chose these geometries in particular so that we could compare our results to the computational and experimental results presented by Lord et al. \cite{lord2010} and the computational results obtained by Liu et al.\cite{liu2019}  Lord et al. \cite{lord2010} fit a five-mode Rolie-Poly model to experimental data of DOW1568 and demonstrated that their computational method, first described by Tenchev et al. \cite{tenchev2008}, was able to qualitatively match principal stress differences measured experimentally along the inflow-outflow center-lines. Liu et al. \cite{liu2019} studied the same model geometries as Lord et al. \cite{lord2010}, but instead utlized a single-mode Rolie-Poly model using only the dominant mode presented by Lord et al. \cite{lord2010}. Nonetheless, they were still able to qualitatively match the principal stress difference inflow-outflow center-line data reported by Lord et al. \cite{lord2010}. In the following results, we use the single-mode Rolie-Poly model used by Liu et al. \cite{liu2019}. These parameters are listed in Table \ref{tab:Params_Rolie_Poly}. Additionally, for both the cross-slot and contraction-expansion slit flows, we decided to limit our scope of study to the narrower three-point B-spline regularized delta function and an $M_{\text{fac}}$ value of $\frac{1}{4}$. A smaller $M_{\text{fac}}$ value of $\frac{1}{4}$ was chosen because we found both the cross-slot and contraction-expansion slit geometries to be much more susceptible to fluid leaks than the cylinder and inclined channel geometries described above. We paired the smaller $M_{\text{fac}}$ value of $\frac{1}{4}$ with the use of the three-point B-spline regularized delta function since the three-point B-spline kernel has a narrower window of support leading to increased interpolation accuracy for functions with jump discontinuities, and is continuously differentiable leading to the suppression of spurious errors when using smaller $M_{\text{fac}}$ values. We note that consistency of the three point B-spline kernel was also pointed out in a recent publication by Lee and Griffith \cite{lee2022} which utilized the immersed-finite-element/difference (IFED) method developed by Griffith and Luo \cite{griffith2017}. \par 

\begin{table}[t!]
\caption{Model parameters for simulating DOW1568 flow through a contraction-expansion slit
geometry and a cross-slot geometry.}
    \centering
	\begin{tabular}{|c|c|c|c|c|c|c|c|}\hline
		$\rho~(\text{g/cm$^3$})$&$\mu~(\text{Pa$\cdot$s})$&$\tau_{\text{d}}~(\text{s})$ & $\tau_{\text{R}}~(\text{s})$&$G~(\text{Pa})$ & $\mu_{\text{p}}~(\text{Pa$\cdot$s})$ &$\beta$ & $\delta$\\\hline
		1.0 & 41.348 & 0.05623 & 0.1 & 72800 & 4093.544 & 0 & -0.5 \\\hline
	\end{tabular}
    \label{tab:Params_Rolie_Poly}
\end{table}
To compare our computational results obtained using the IB method to the experimental and computational studies we compute the principal stress difference (PSD),$\sqrt{\left(\sigma_{\text{p}_{xx}} - \sigma_{\text{p}_{yy}}\right)^2 + 4\sigma_{\text{p}_{xy}}^2}$, along the inflow outflow center-lines of each geometry. 
The PSD is a commonly reported experimental quantity since the PSD is linearly proportional to the flow induced birefringence which can be readily measured by monitoring the polarization state of a light ray passing through the polymer solution \cite{clemeur2004}.
\subsubsection{Rolie-Poly Flow through a Contraction Expansion Slit}
\begin{figure}[t!]
    \centering
    \includegraphics[scale = 0.6]{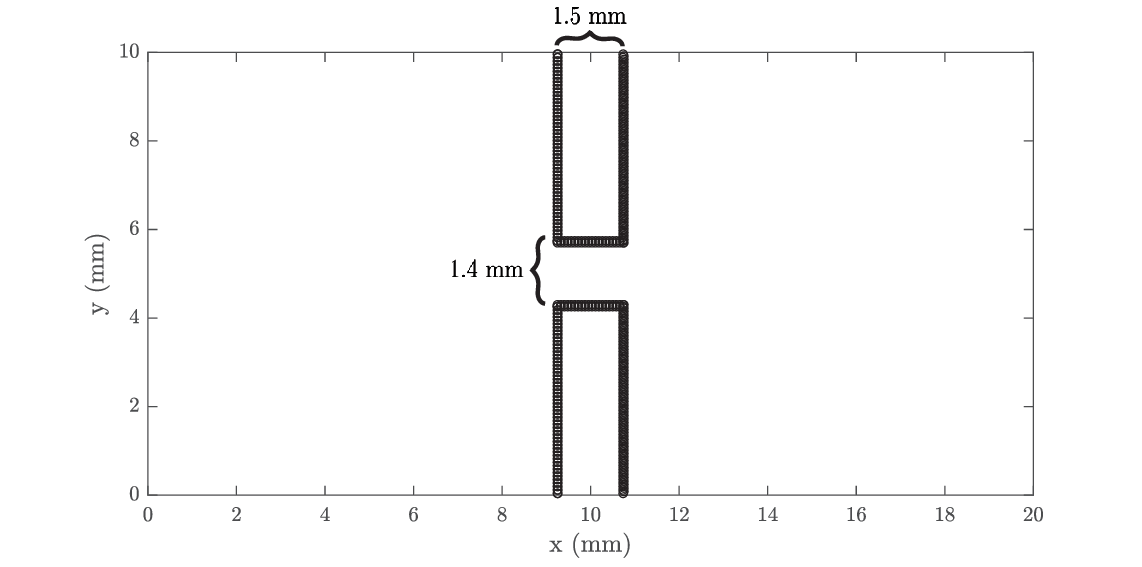}
    \caption{Illustration of the contraction expansion slit geometry with $h = \frac{W}{32}$. }
    \label{fig:Contraction_Slit_geom}
\end{figure}

As shown in Figure \ref{fig:Contraction_Slit_geom}, the contraction expansion slit geometry is modeled on a two dimensional rectangular channel of length $L= 20$ mm and width $W = 10$ mm. The contraction is generated by two rectangular blocks which adhere to the physical domain with a width of 1.5 mm and slit depth of 1.4 mm yielding a contraction ratio of roughly 7.14:1.  
\par
\begin{figure}[t!]
    \includegraphics[width = 0.95\linewidth, left]{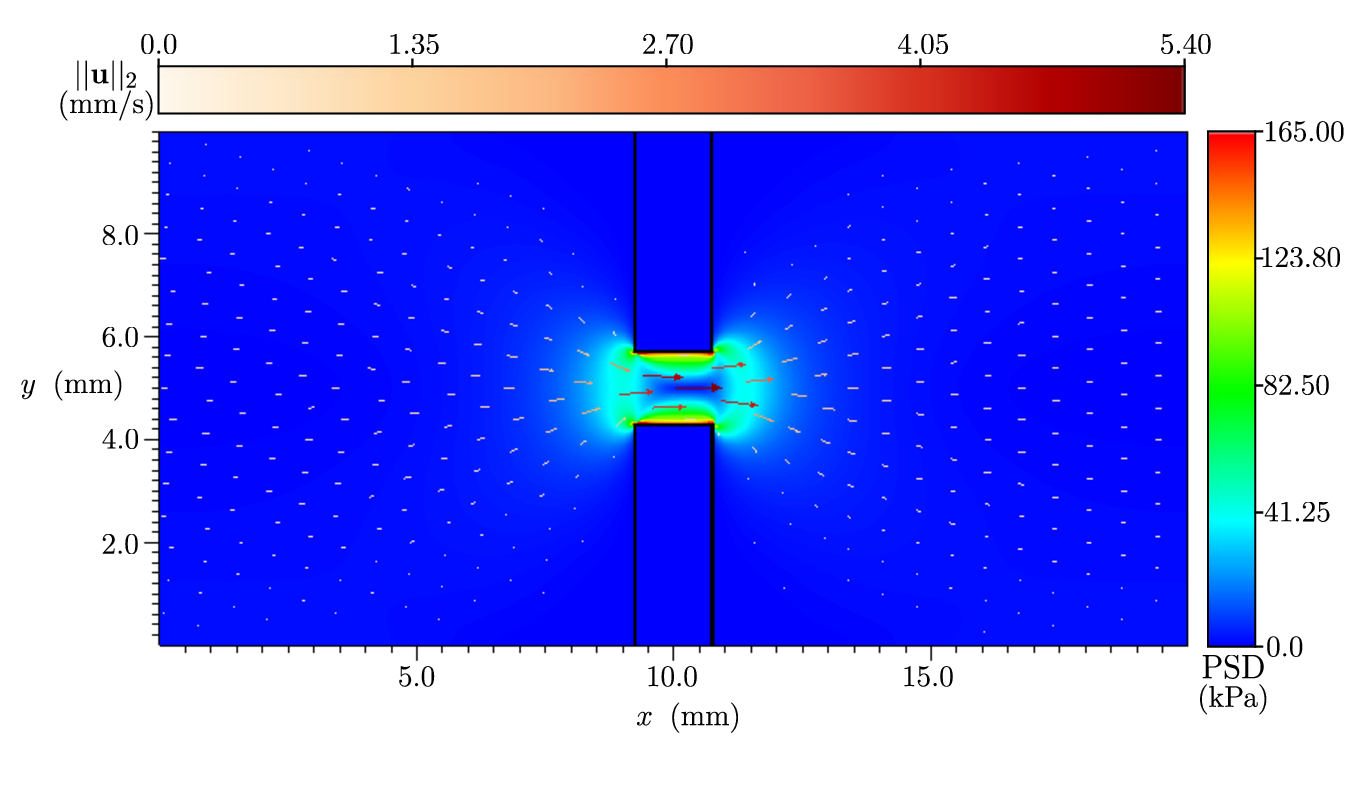}
    \caption{Visualization of steady 2D Rolie-Poly flow through a contraction-expansion slit. The IB markers outlining the Lagrangian blocks are displayed in black. The velocity vector field is illustrated atop the principal stress difference field in red and white with vectors shaded in darker hues of red corresponding to larger velocity magnitudes. The uniform Cartesian grid is spaced with a uniform grid increment of $h = \frac{W}{512}$. The Lagrangian markers outlining the channel are spaced with an arc length increment of $\frac{h}{4}$ to prevent fluid leaks. The physical parameters used for the simulation are detailed in Table \ref{tab:Params_Rolie_Poly}.}
    \label{fig:Flow_Through_Contraction_Slit}
\end{figure}

Although there is no analytic solution to the steady-state Rolie-Poly channel flow, we specify for the inflow boundary condition a parabolic normal velocity profile and a zero tangential velocity 
\begin{equation}
    u_{\text{inflow}} = 0.78\cdot\frac{y(10-y)}{25}\;\unit[per-mode = symbol]
{\mm\per\second}, \quad v_{\text{inflow}} = 0\;\unit[per-mode = symbol]
{\mm\per\second}.
\end{equation}
For the outflow boundary condition, we employed a zero normal traction and a zero tangential velocity. No slip conditions are utilized for the velocities along the walls confining the channel and the immersed boundary. For the components of the conformation tensor, we specify homogeneous Neumann boundary conditions along each boundary of the computational domain. While this procedure is not correct even for fully developed flow conditions, the errors introduced as a result remain localized near the boundary \cite{alves2021}. We set the initial conditions to be at rest with $\velocity = \mathbf{0}$ and $\conformTens = \mathbb{I}$. Figure \ref{fig:Flow_Through_Contraction_Slit} shows the down-sampled computed velocity vector field $\mathbf{u}$ atop the computed principal stress difference contours for the finest discretization used. 
We perform a grid convergence study with three different uniform spatial discretizations $(N_x,N_y) = (256,128), (512,256)$ and $(1024,512)$ so that the uniform spaced Cartesian grid is obtained with an increment of $h = \frac{W}{N_y}$, correspondingly. We choose $\Delta t = 1.075\times 10^{-5}/N_y$ $\si{\s}$. We set the penalty spring parameter $\kappa$ using equation \eqref{eq:scaling_penalty} with a dimensionless constant $C = 2.5$. Doing so ensures that Lagrangian markers move no further than $\frac{h}{2}$ from their initial locations throughout the simulation. Table \ref{tab:convergence_rates_contraction_expansion_slit} reports empirical rates of convergence in the components of the conformation tensor $\conformTens$ and in the components of the velocity $\mathbf{u}$. The empirical convergence rates were computed using Richardson extrapolation. 

\begin{table}[t!]
\caption{Global convergence rates for the components of the velocity, pressure, and stress tensor for two dimensional Rolie-Poly flow though a contraction-expansion slit. The physical parameters used in the simulation are given in Table \ref{tab:Params_Rolie_Poly}. The contraction expansion slit is represented using Lagrangian markers spaced $\frac{1}{4}h$ apart from one another to prevent fluid leaks. Empirical rates of convergence were computed using Richardson extrapolation on uniform Cartesian grids with grid increment sizes of $h = \frac{W}{128},\frac{W}{256},$ and $\frac{W}{512}$.}
    \centering
    \begin{tabular}{c|c|c}
         & $\sigma_{\text{p}_{xx}}$ & $\sigma_{\text{p}_{xy}}$  \\\hline
        $L^1$ rate & 0.8962 & 0.9498 \\
        $L^2$ rate & 0.3645 & 0.5555 \\
        $L^{\infty}$ rate & -0.6709 & -0.3123 \\\hline
         & $\sigma_{\text{p}_{yy}}$ & $\mathbf{u}$ \\\hline
        $L^1$ rate & 1.0189 & 1.1877 \\
        $L^2$ rate & 0.3122 & 1.2517 \\
        $L^{\infty}$ rate & -1.1709 & 0.6790 \\\hline
    \end{tabular}
    \label{tab:convergence_rates_contraction_expansion_slit}
\end{table}
Similar to the convergence properties observed for Oldroyd-B flow past a cylinder, we found the components of the polymeric stress tensor $\polymerStress$ to converge at a first order rate in $L^1$, about a half order rate in $L^2$, and fails to converge pointwise. However, the velocity $\mathbf{u}$ appears to be converging pointwise at rate less than first order. This observed reduced rate of convergence is likely due to the presence of the sharp corners in our model geometry. Recall that in the proof of the convergence of the immersed boundary method for Stokes flow, pointwise error estimates were obtained under the assumption that the immersed boundary was $C^2$-regular \cite{mori2008}.\par
\begin{figure}[t!]
\centering
\includegraphics[width = 0.9\textwidth]{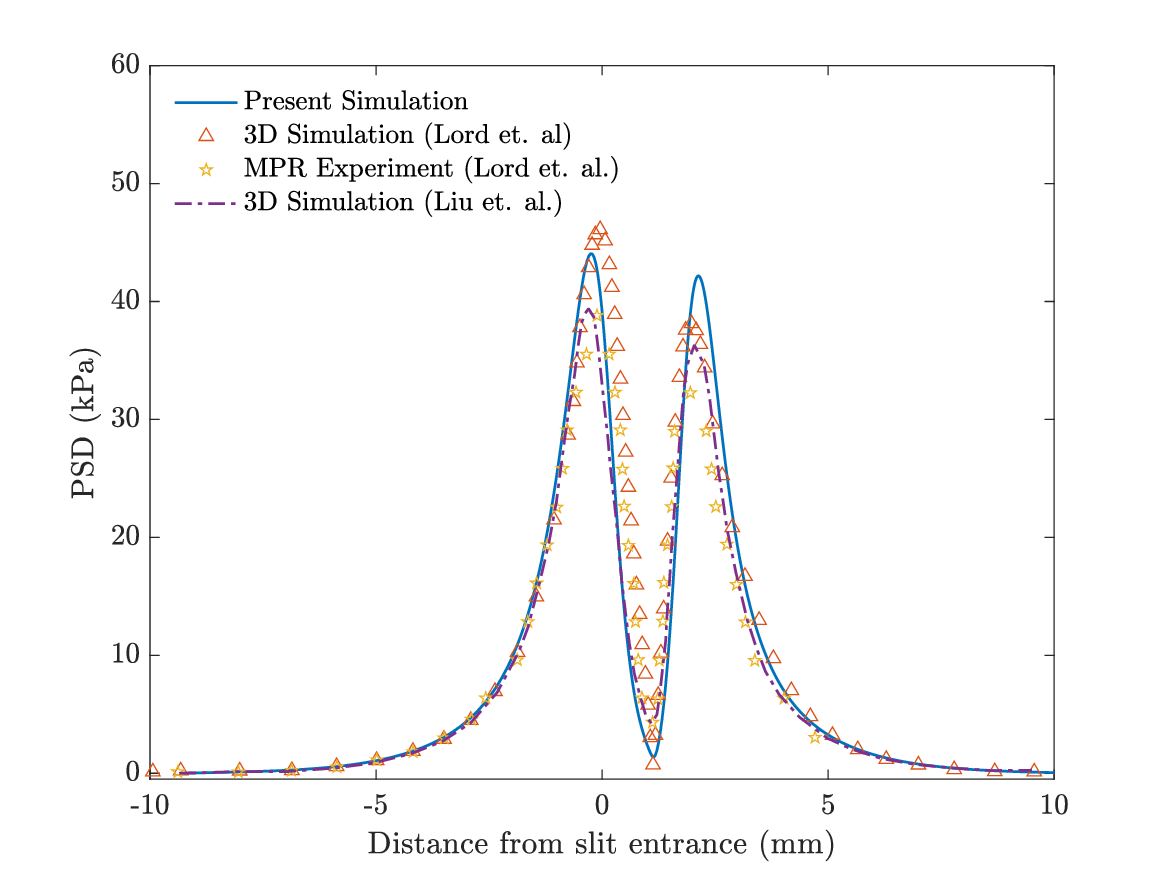}
\caption{Reported computational and experimental PSD results for mono-disperse polystyrene along the inflow-outflow centerline of the contraction-expansion slit geometry. The mono-disperse polysyrene solution was modeled using the Rolie-Poly constitutive equation. The simulated PSD reported herein using the IB method is illustrated by the blue solid line. The simulated PSD reported by Liu et al. \cite{liu2019} and Lord et al. \cite{lord2010} are displayed using the purple dashed line and the orange triangle markers, respectively. The gold stars highlight the experimental PSD reported by Lord et al. The IB method slightly over predicts the PSD near the slit exit, but appears to appears to qualitatively match the previously reported PSD profiles.}
\label{fig:PSD_centerline_comparison_contraction_expansion_slit}
\end{figure}

    
In Figure \ref{fig:PSD_centerline_comparison_contraction_expansion_slit}, we compare our computed principal stress difference and velocities on the  the finest grid $(N_x,N_y) = (1024,512)$ to the three dimensional simulations and experimental data presented by Lord et al.\cite{lord2010} and Liu et al. \cite{liu2019}. Qualitatively, the PSD computed by the IB method along the inflow-outflow centerline agrees quite well with the previous computational and experimental results; however, it appears that the IB method slightly over-predicts the PSD at the slit exit. At steady-state, the differences in pressure measured at the inlet and outlet of the contraction is approximately 2.2 bars. This finding is in good agreement with the 2.0 bar difference reported by Lord et al. \cite{lord2010}

Additionally, we also note that along the inflow-outflow centerline, we observe an empirical pointwise convergence rate of 1.69 for the PSD indicating that, away from immersed boundaries, pointwise convergence in the components of the extra-stress tensor is achieved. The empirical pointwise convergence rate for the velocity along the inflow-outflow centerline is $1.29$. 

\subsubsection{Rolie-Poly Flow through a Cross-Slot}
\begin{figure}[t!]
    \centering
    \includegraphics[width = 0.75\textwidth]{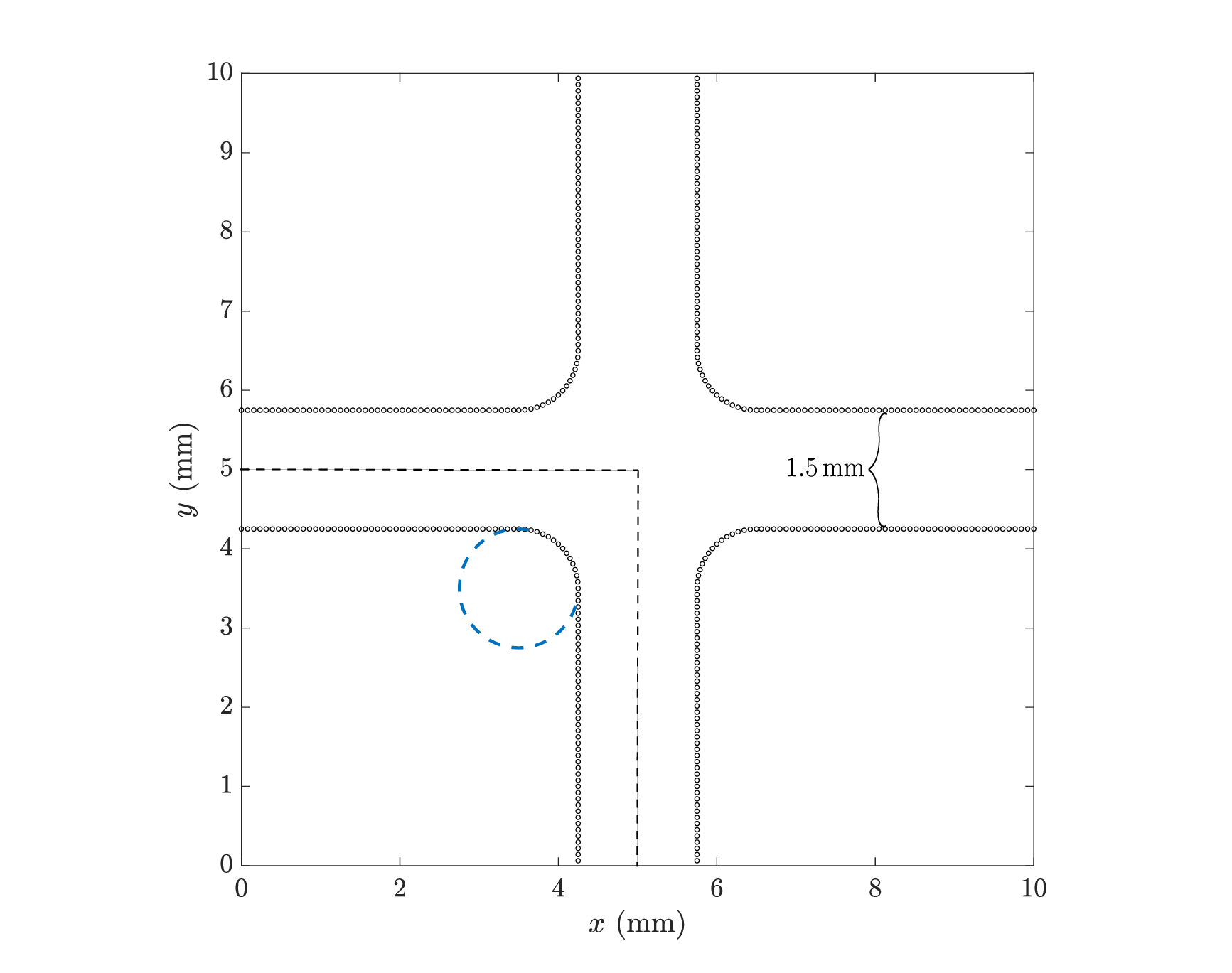}
    \caption{Illustration of the cross-slot geometry when $h = 0.3125$ mm. The inflow-outflow centerline is represented by the dashed black line. The circle used to generate the curvature of the cross-slot walls is illustrated by the dashed blue line.}
    \label{fig:Cross_Slot_Geom}
\end{figure}

We next investigate the capability of the IB-method to simulate Rolie-Poly flow through a cross-slot geometry. The cross-slot geometry, illustrated in Figure \ref{fig:Cross_Slot_Geom}, consists of four curved walls embedded in a square channel of length $L = 10$ mm. The width of the inlets and outlets is $1.5$ mm. The curvature of the walls is generated by placing the Lagrangian markers along a circle of radius $0.75$ mm which connects points extending from the inlet to points extending to the outlet.

Like the contraction-expansion slit example above, we employed homogeneous Neumann boundary conditions throughout the computational domain for the components of the conformation tensor. For the components of the velocity, the inflow boundary conditions for $4.25 \text{mm} \leq y \le 5.75 \text{mm}$ are given by
\begin{align}
&    u_{\text{left}} =
     3.15\left(1 - \frac{(y-5.0)^2}{(0.75)^2}\right)\;\unit[per-mode = symbol]
{\mm\per\second} \\ 
&    u_{\text{right}} = -3.15\left(1 - \frac{(y-5.0)^2}{(0.75)^2}\right)\;\unit[per-mode = symbol]
{\mm\per\second} \\
& v_{\text{left}} = v_{\text{right}} = 0\;\unit[per-mode = symbol]
{\mm\per\second}.
\end{align}
While for $5.25 \text{mm} < y < 4.25 \text{mm}$ the tangential velocity and normal homogenous traction were set to zero. At the outlets, homogenous traction boundary conditions are employed in the normal direction and homogenous Dirichlet boundary conditions used for the tangential component of the velocity.  \par
Similar to the analysis presented in the contraction-expansion slit example above, we perform a grid convergence study using three different uniform spatial discretizations of the 10 mm $\times$ 10 mm domain the cross-slot is embedded in. The spatial discretizations used are $(N_x,N_y) = (N,N) = (128,128), (256,256),$ and $(512,512)$ so that the uniform mesh width is $h = L/N$. For each discretization, we once again apply a time step size of = $\Delta t = 1.075\times10^{-5}/N$~s and the penalty parameter is set according to equation \eqref{eq:scaling_penalty} with a dimensionless constant $C = 2.5$. Table \ref{tab:conv_rates_cross_slot} reports the empirical global convergence rates.
\begin{table}[t!]
\caption{Global convergence rates for the components of the velocity and polymeric stress tensor for Rolie-Poly flow through the cross-slot geometry. Rates of convergence were estimated using Richardson extrapolation with solution components computed on increasingly fine uniform Cartesian grids with grid increments of $h = \frac{L}{128},\frac{L}{256},$ and $\frac{L}{512}$. The physical parameters used in each simulation are detailed in Table \ref{tab:Params_Rolie_Poly}. The Lagrangian markers were placed with approximate arc length increments of $\frac{1}{4}h$ to prevent fluid leaks through the structure.}
    \centering
    \begin{tabular}{c|c|c}
         & $\sigma_{\text{p}_{xx}}$ & $\sigma_{\text{p}_{xy}}$  \\\hline
        $L^1$ rate & 1.541 & 1.016 \\
        $L^2$ rate & 0.937 & 0.554  \\
        $L^{\infty}$ rate & 0.0380 & 0.232 \\\hline
         & $\sigma_{\text{p}_{yy}}$ & $\mathbf{u}$ \\\hline
        $L^1$ rate & 1.641 & 2.365 \\
        $L^2$ rate & 1.011 & 1.967 \\
        $L^{\infty}$ rate & -0.4294 & 1.674 \\\hline
    \end{tabular}
    \label{tab:conv_rates_cross_slot}
\end{table}
We reiterate that the global convergence rates of the polymeric stress tensor exhibit convergence in the $L^1$ and $L^2$ grid norms. Notice also that these global convergence rates are in generally higher for the cross-slot geometry than for the contraction-expansion slit geometry. This is likely a consequence of the cross-slot geometry's smooth parameterization, whereas the parameterization of the contraction-expansion slit geometry is only piece-wise smooth and includes sharp corners. \par
\begin{figure}[h!]
\centering
\includegraphics[width = 0.9\textwidth]{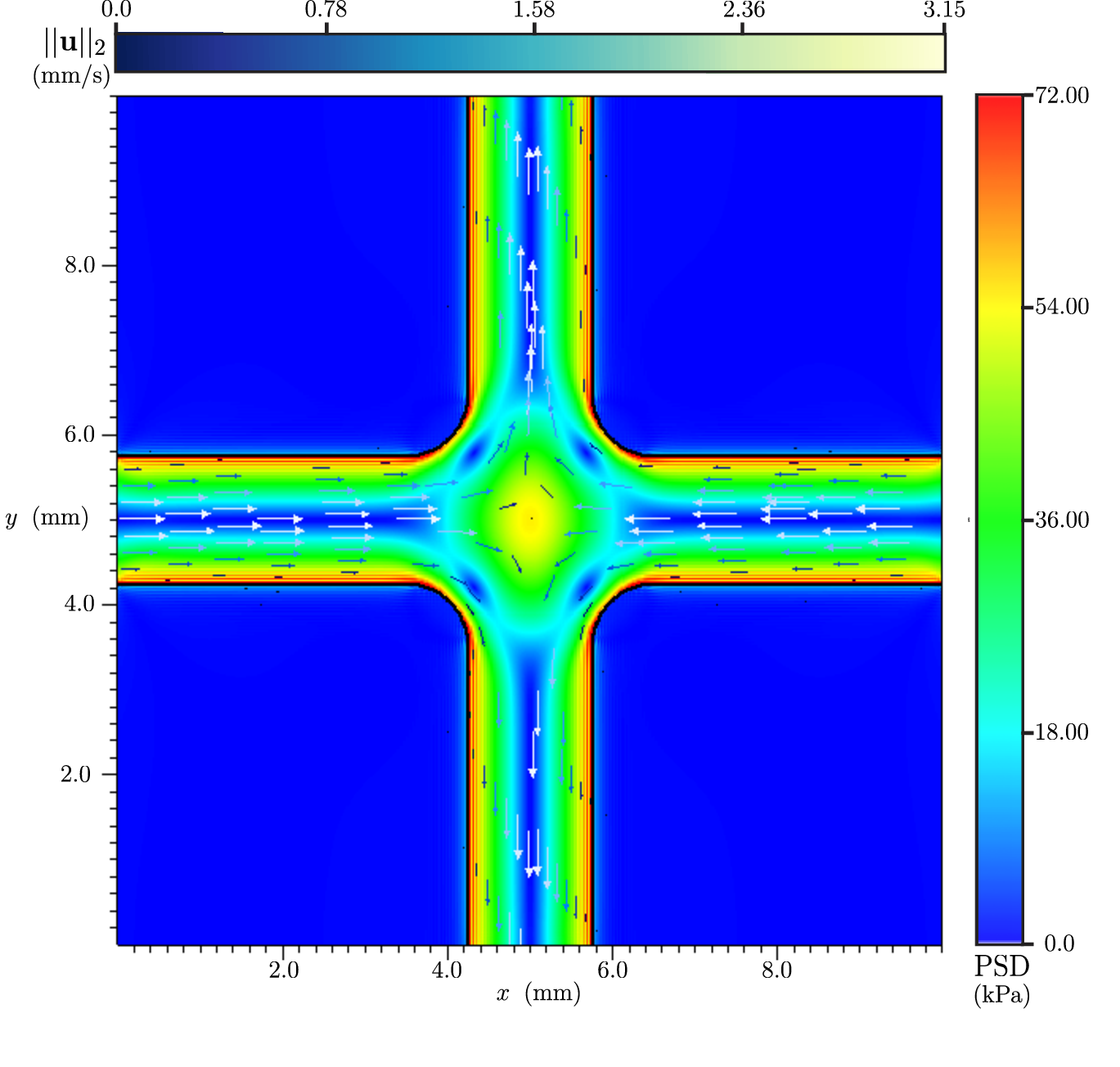}
\caption{Visulization of steady 2D Rolie-Poly flow through the cross-slot geometry. The velocity vector field is illustrated atop the principal stress difference field in blue with lighter hues corresponding to larger velocity magnitudes. The largest stresses are situated along the boundary and at the stagnation point located in the center of the cross-slot geometry. The physical parameters used in this simulation are detailed in Table \ref{tab:Params_Rolie_Poly}.}
\label{fig:Cross_Slot_SS}
\end{figure}
\begin{figure}[!htb]
\centering
\includegraphics[width = 0.9\textwidth]{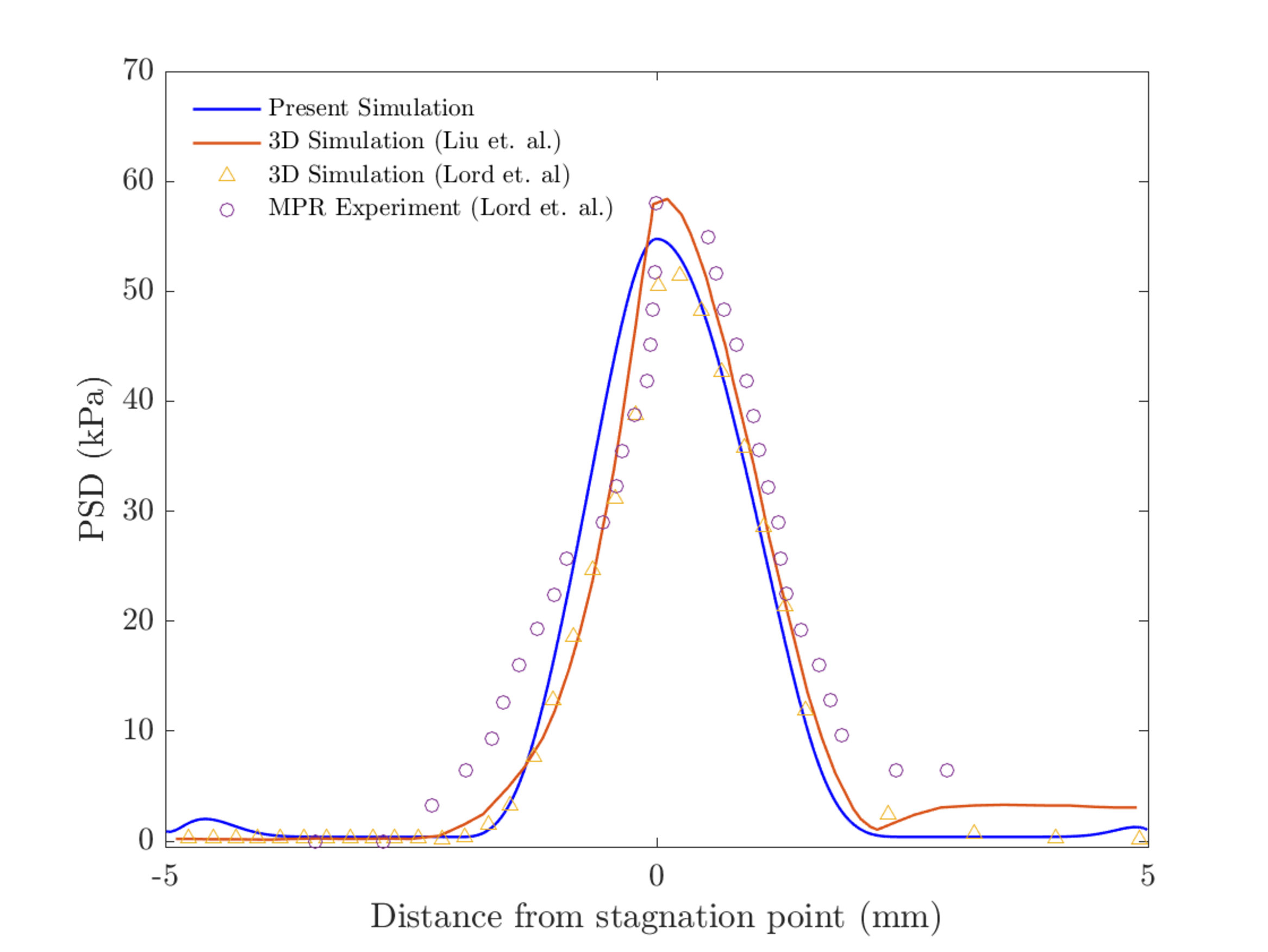}
\caption{Comparison of the computed and experimental PSD profiles of mono-disperse polystyrene along the inflow-outflow centerline associated with the cross-slot geometry. Computational PSD profiles obtained by Lord et al. (yellow triangles) utilized a five-mode Rolie-Poly model while the present profile (blue line) and profile reported by Liu et al. \cite{liu2019} (red line) utilized a single-mode Rolie-Poly model associated with the dominant Rolie-Poly relaxation mode reported by Lord and coworkers \cite{lord2010} whose parameters are listed in Table \ref{tab:Params_Rolie_Poly}. The IB method compares best the simulated PSD profile reported by Lord et al. The bumps present in PSD profile obtained using the IB method are due to the prescription of incorrect boundary conditions at the inlet and outlet of the cross slot.}
\label{fig:Cross_Slot_PSD_comparison}
\end{figure}

 Figure \ref{fig:Cross_Slot_SS} presents the global steady-state PSD for the finest uniform discretization employed. We highlight in particular the observed PSD asymmetry located along the inflow-outflow centerline which been well documented in a number of experimental and computational studies \cite{auhl2011,mackley2011,lord2010,castillosanchez2022}. 
In Figure \ref{fig:Cross_Slot_PSD_comparison}, we compare our computed principal stress differences along the inflow-outflow centerline to the principal stress differences computed and experimentally obtained by Liu et al. \cite{liu2019} and Lord et al.\cite{lord2010}. Overall, the IB method qualitatively reproduces the computed and experimentally observed principal stress difference.
We also point out the presence of small bumps in the values of principal stress difference located near the inflow-outflow boundaries which are not present in the results presented by Liu et al. and Lord et al. These bumps are not a numerical artifact of the IB method but instead result from to the prescription of the homogenous Neumann boundary conditions for the conformation tensor along the inflow and outflow boundaries. Numerical simulations of channel flow with a Rolie-Poly model suggest the presence of these bumps are a result of the prescribed boundary conditions being incorrect. The pseudo-periodic boundary conditions employed by Tenchev et al. \cite{tenchev2011} greatly reduces these numerical artifacts.

\section{Discussion and Conclusion}
The numerical experiments conducted herein, using both the idealized Oldroyd-B model and the statistical physics-derived Rolie Poly model, demonstrate that the IB method can be successfully and accurately applied to the investigation of viscoleastic flows around and through immersed structures. The IB method is able to provide globally convergent values of the conformation and stress tensor in the $L^1$ and $L^2$ grid norms. Although fluid stresses and values of the polymeric stress generated by the IB method do not converge pointwise along the interface, the methodology nonetheless generates convergent net forces, such as drag, that are computed by integrating over the Lagrangian forces defined along the immersed boundary. A key finding of this study is that the net forces computed using the IB method appear to converge at first order, analogous to the convergence rates of net forces obtained using traditional finite-element discretizations of viscoelastic flow past stationary structures \cite{tenchev2011}. Furthermore, as the Rolie-Poly contraction-expansion slit and cross-slot flow examples indicate, the IB method can qualitatively predict flow induced birefringence measurements. Therefore, the IB method, coupled with its ease of implementation, is competitive for investigating the predictive capability of a viscoelastic constitutive equation. \par 

Simulations of Oldroyd-B flows through a inclined channel and past a cylinder examples reveal that the IB method is generally more accurate when it uses a smooth and narrowly supported regularized delta function. The smoothness of the regularized delta function helps to cull spurious oscillations which may arise if the $M_{\text{fac}}$ value employed is too small. For instance, as illustrated in Table \ref{tab:Drag_Errors}, the three-point B-spline regularized delta function, which is continuously differentiable, uniformly generates accurate drag coefficients for each of the $M_{\text{fac}}$ values tested. Conversely, the piecewise-linear regularized delta function, which is only continuous, produces comparatively accurate values of the drag coefficient only when paired with the coarsest $M_{\text{fac}} = 2$ setting. We remark that for the finest grid resolution tested $\left(h_{\text{finest}} = \frac{H}{256}\right)$, using the piecewise-linear regularized delta function with $M_{\text{fac}} = 2$ or the three-point B-spline regularized delta function at any of the $M_{\text{fac}}$ values tested, results in drag coefficients whose relative errors are on the same order of magnitude as those computed using the finite element discretization described by Tenchev et al. \cite{tenchev2011} for a comparably resolved mesh. However, in general, we find that the IB method underestimates the polymeric stress near the boundary of the cylinder with smoother and more broadly supported regularized delta functions providing the smallest estimates.

We note that, as mentioned above in the Rolie-Poly flow examples, in which relatively larger pressure loads are present, the $M_{\text{fac}}$ value  needs to be made small enough in order to prevent fluid leaks through the structure. In these cases, the use of a relatively smooth but narrowly supported regularized delta function such as the three-point B-spline or the three-point IB regularized delta function \cite{roma1999} that are both continuously differentiable is suggested.  The property of narrow support improves the interpolation accuracy of the regularized delta function when applied to functions with discontinuities in their derivatives. Because the B-spline family of functions may be characterized by their minimal support property\cite{schoenberg1964}, we recommend the use of the B-spline family of regularized delta functions when implementing the IB method to solve for viscoelastic flow dynamics. 

Some of our findings concerning how the $M_{\text{fac}}$ parameter and choice of regularized delta function impacts the accuracy of the IB method are similar to those reported by Lee and Griffith whose recent study provides a much broader analysis of the regularized delta function and $M_{\text{fac}}$ value pairing, but in the context of the IFED method \cite{lee2022}. For example, Lee and Griffith found that, in general, utilizing a regularized delta function of narrower support generally provides more accurate solutions of flows past rigid bodies. We believe this finding can be rationalized, at least heuristically, by observing that in one dimension, regularized delta functions with smaller mean values on the half-line provide more accurate interpolants of functions whose derivatives have a discontinuity at the interpolation point of interest. Additionally, we also find that when pressure loads on the immersed structure are substantial, smaller $M_{\text{fac}}$ values must be employed to prevent fluid leaks through the structure. However, we do observe some differences from Lee and Griffith. For example, so long as pressure loads on the structure are mild, Lee and Griffith found that larger values of the $M_{\text{fac}}$ parameter tend to lead to both more accurate Lagrangian forces and velocities \cite{lee2022}. However, we find that the $M_{\text{fac}}$ parameter has little influence over the accuracy of the computed velocity, and only affects the accuracy of the Lagrangian forces when the regularized delta function utilized has narrow support and is less regular. For example, the drag coefficient computed using the piecewise linear regularized delta function provides both the most accurate estimate of the drag coefficient when $M_{\text{fac}} = 2$ and the least accurate estimate of the drag coefficient when $M_{\text{fac}} = \frac{1}{2}$. The increase in error for smaller values of $M_{\text{fac}}$ appears to be caused by non-physical spurious oscillations in the Lagrangian positions and forces, which appear to be more prominent if using regularized delta functions with less regularity and narrower support paired with a small $M_{\text{fac}}$ value. Yang et al.~noted that utilizing smoother regularized delta functions helps to limit the spurious oscillations present in the Lagrangian forces \cite{yang2009}. However, we believe the presence of these spurious oscillations to be engendered not only by the lack of regularity associated with a regularized delta function, but also the $M_{\text{fac}}$ value employed. \par 

In closing, we reiterate that the IB method, with its ease of implementation, is an effective strategy for simulating viscoelastic flow through and around complex geometries. Since the IB method obtains convergent values of the stress away from immersed boundaries, the IB method may be paired with flow-induced birefringence experiments to evaluate the accuracy and predictive power of viscoelastic constitutive equations. For scenarios involving stationary bodies, we advise employing a regularized delta function that is both narrow and smooth. This approach is particularly beneficial for coarser grid discretizations, leading to more precise flow simulations and calculations of Lagrangian forces. Moreover, in situations where the pressure loads on structures are not substantial enough to induce leaking, it is preferable to opt for a Lagrangian mesh which is relatively coarse compared to the background Cartesian grid. Doing so helps minimize spurious oscillations in the Lagrangian forces, thus enhancing their accuracy.

While this study primarily addresses stationary structures at relatively low Reynolds and Weissenberg numbers, preliminary evidence suggests that the insights and recommendations garnered here should be broadly applicable to a diverse array of viscoelastic flow scenarios with stationary structures. This includes flows characterized by either high Reynolds or high Weissenberg numbers. However, it's important to note that when managing high Weissenberg number flows, one often needs to evolve either the square-root\cite{balci2011} or logarithm\cite{fattal2004, fattal2005} of the conformation tensor to maintain numerical stability and improve accuracy. 

In the future, we aim to assess the effectiveness of the IB method applied to viscoelastic flows involving moving structures. However, the current lack of robust benchmark studies in this area presents a significant challenge. The development of such benchmark studies should be informed by, and measured against, rigorously obtained experimental data. One promising approach is the modeling of micro-sphere rheology experiments for which there is a large amount of experimental data available for viscoelastic fluids with a variety of characteristics. The establishment of these benchmark studies is crucial, as they will provide the scientific community with validated frameworks to accurately predict and analyze viscoelastic fluid strucuture interaction, ultimately advancing applications in various fields ranging from medicine to manufacturing.

\FloatBarrier

\section{Acknowledgments}
Cole Gruninger is grateful for support from the Department of Defense (DoD) through the National Defense Science and Engineering Graduate (NDSEG) Fellowship Program. Boyce E. Griffith thanks the National Science Foundation (NSF) for support under grant numbers NSF DMS 1410873, NSF DMS 1664645, NSF OAC 1450327, and NSF OAC 1931516. Aaron Barret gratefully acknowledges support from the National Institutes of Health for support under grant numbers NIH U01HL143336 and NIH R01HL157631. M. Gregory Forest acknowledges support from the NSF under grant number NSF CISE-1931516 and the Alfred P. Sloan foundation. 

\renewcommand\refname{REFERENCES}
\bibliographystyle{unsrt}
\bibliography{refs}

\end{document}